\documentclass[aps,prb,twocolumn,superscriptaddress]{revtex4}

\usepackage{graphics}
\usepackage{epsfig}
\usepackage{graphicx}
\usepackage{dcolumn}
\usepackage{tabularx}
\usepackage{bm}
\usepackage{graphicx}
\usepackage{dcolumn}
\usepackage{rotating}
\usepackage{supertabular}
\usepackage{amssymb}
\usepackage{amsmath,theorem}
\usepackage{threeparttable}
\usepackage{multirow}
\usepackage{units}
\usepackage{color}
\newlength{\myleno}
\newlength{\mylent}
\newcommand{\fr}{\ensuremath{\mathbf{r}}}
\newcommand{\fx}{\ensuremath{\mathbf{x}}}

\newcommand{\LRPBE}{{LC-$\omega$PBE}}

\newcommand{\cRPA}{{RPA}}
\newcommand{\cSE}{{SE}}
\newcommand{\cRSE}{{rSE}}
\newcommand{\cSOSEX}{RPA+SOSEX}
\newcommand{\ccSOSEX}{SOSEX}
\newcommand{\RPA}{{(EX+\cRPA)@PBE}}
\newcommand{\RPAh}{{HF+\cRPA @PBE}}
\newcommand{\RPASE}{{(EX+\cRPA+\cSE)@PBE}}
\newcommand{\RPARSE}{{(EX+\cRPA+\cRSE)@PBE}}
\newcommand{\SOSEX}{{(EX+\cSOSEX)@PBE}}
\newcommand{\SOSEXh}{{HF+(\cSOSEX)@PBE}}
\newcommand{\SOSEXRSE}{{(EX+\cSOSEX+\cRSE)@PBE}}

\newcommand{\ket}[1]{\ensuremath{ \vert #1 \rangle }}
\newcommand{\braket}[2]{\left\langle \, #1 \, | \, #2 \right \rangle}

\newcommand{\A}{\mathbf{A}}
\newcommand{\B}{\mathbf{B}}
\newcommand{\T}{\mathbf{T}}

\newcommand{\GDV}{\textsc{gaussian}}
\newcommand{\AIMS}{\textsc{FHI}-aims}
\newcommand{\VASP}{\textsc{vasp}}
\newcommand{\eps}{\varepsilon}

\begin{document}

\title{Assessment of correlation energies based on the random-phase approximation}

\author{Joachim Paier}
\affiliation{Institut f{\"u}r Chemie, Humboldt-Universit{\"a}t zu Berlin, Unter den Linden 6, 10099 Berlin, Germany}

\author{Xinguo Ren}
\affiliation{Fritz-Haber-Institut der Max-Planck-Gesellschaft, Faradayweg 4-6, D-14195 Berlin, Germany}
\affiliation{European Theoretical Spectroscopy Facility (ETSF)}

\author{Patrick Rinke}
\affiliation{Fritz-Haber-Institut der Max-Planck-Gesellschaft, Faradayweg 4-6, D-14195 Berlin, Germany}
\affiliation{European Theoretical Spectroscopy Facility (ETSF)}

\author{Gustavo E. Scuseria}
\affiliation{Department of Chemistry, Rice University, Houston, Texas 77005, USA}
\affiliation{Department of Physics and Astronomy, Rice University, Houston, Texas 77005, USA}

\author{Andreas Gr{\"u}neis}
\affiliation{Faculty of Physics and Center for Computational Materials Science, Universit{\"a}t Wien, Sensengasse 8/12, A-1090 Wien, Austria}

\author{Georg Kresse}
\affiliation{Faculty of Physics and Center for Computational Materials Science, Universit{\"a}t Wien, Sensengasse 8/12, A-1090 Wien, Austria}

\author{Matthias Scheffler}
\affiliation{Fritz-Haber-Institut der Max-Planck-Gesellschaft, Faradayweg 4-6, D-14195 Berlin, Germany}
\affiliation{European Theoretical Spectroscopy Facility (ETSF)}

\date{\today}

\begin{abstract}
The random-phase approximation to the ground state correlation energy (\cRPA) in combination with exact exchange (EX) 
has brought Kohn-Sham (KS) density functional theory one step closer towards a universal, 
``general purpose first principles method''.
In an effort to systematically assess the influence of several correlation energy contributions 
beyond RPA, this work
presents dissociation energies of small molecules and solids, activation energies 
for hydrogen transfer 
and non-hydrogen transfer reactions, as well as reaction energies for a number of common test sets.
We benchmark EX+\cRPA\ and several flavors of energy functionals going beyond it:  second-order screened exchange (SOSEX), single 
excitation (SE) corrections, renormalized single excitation (\cRSE) corrections, as well as their 
combinations. 
Both the single excitation correction as well as the SOSEX contribution to the correlation energy significantly 
improve upon the notorious tendency of EX+\cRPA\ to underbind.
Surprisingly, activation energies obtained using EX+\cRPA\ based on a KS reference alone are remarkably accurate.
RPA+SOSEX+\cRSE\ provides an equal level of accuracy for reaction as well as activation energies 
 and overall gives the most balanced performance, which makes it applicable to a wide range of 
systems and chemical reactions.
\end{abstract}


\maketitle
%
%

\section{Introduction\label{sec:intro}}
In the context of first principles electronic structure theory 
``exact-exchange plus correlation in the random-phase approximation (EX+RPA)" 
\cite{Bohm/Pines:1953,Gell-Mann/Brueckner:1957} has recently generated renewed and widespread 
interest.~\cite{Furche:2001,Fuchs/Gonze:2002,furche:jcp:05,Scuseria/Henderson/Sorensen:2008,Janesko/Henderson/Scuseria:2009,Toulouse/etal:2009,paier:jcp:10,Marini/Gonzalez/Rubio:2006,Jiang/Engel:2007,Harl/Kresse:2008,Harl/Kresse:2009,Lu/Li/Rocca/Galli:2009,Dobson/Wang:1999,Rohlfing/Bredow:2008,Ren/Rinke/Scheffler:2009,Schimka/etal:2010,Zhu/etal:2010,Lebegue/etal:2010,eshuis:jcp:2010,Ismail-Beigi:2010,goltl:jcp:2011,eshuis:jpcl:2011,ren:prl:11,Hesselmann/Goerling:2011,Hesselmann/Goerling:2011_2,ruzsinszky:jcp:2011,hesselmann:jcp:11,klopper:cpl:11} 
%
In practice, the RPA calculations are most often carried 
 out in a non-self-consistent manner where the exchange-correlation
 (xc) energy contributions are evaluated with input orbitals
 corresponding to an approximate, usually semilocal xc 
 energy functional.
%
The great interest in EX+RPA is largely due to its three attractive features:
%
({\it i}) The exact-exchange energy (EX) cancels the spurious self-interaction error present in the Hartree energy,
({\it ii}) the RPA correlation energy is fully non-local and includes long-range van der Waals (vdW) interactions 
automatically and seamlessly,~\cite{Dobson:1994}
and ({\it iii}) EX+\cRPA\ is applicable to small-gap or metallic systems
by summing up  the sequence of ``ring" diagrams to infinite order. 
The latter is in contrast to order-by-order perturbation theories [e.g. 2nd-order M{\o}ller-Plesset (MP2) \cite{Moller/Plesset:1934}]
which break down for systems with zero gap.
Moreover one can interpret the RPA as an approach that screens the non-local exchange
resulting in a  frequency dependent non-local screened exchange interaction, 
as opposed to conventional or global hybrid functionals where the parameters that reduce 
or ``screen"  the exact-exchange contribution are fixed and system independent.\cite{BHH,B3,PBE14}
Such a system independent ``screening'' is expected to be unreliable for metals or
wide gap insulators, where non-local exchange is almost entirely screened (metals) 
or prevails to a large extent (insulators).

While a critical assessment of EX+\cRPA\ is emerging \cite{furche:jcp:05,Scuseria/Henderson/Sorensen:2008,Janesko/Henderson/Scuseria:2009,Toulouse/etal:2009,paier:jcp:10,Marini/Gonzalez/Rubio:2006,Harl/Kresse:2008,Harl/Kresse:2009,Lu/Li/Rocca/Galli:2009,Dobson/Wang:1999,Rohlfing/Bredow:2008,Ren/Rinke/Scheffler:2009,Schimka/etal:2010,Zhu/etal:2010,Lebegue/etal:2010,eshuis:jcp:2010,Ismail-Beigi:2010,goltl:jcp:2011,eshuis:jpcl:2011,ren:prl:11,Hesselmann/Goerling:2011,ruzsinszky:jcp:2011} some shortcomings have been known for a while. 
Total energies are typically significantly overestimated,\cite{freeman:prb:77,Furche:2001,Jiang/Engel:2007,grueneis:jcp:09,paier:jcp:10,ruzsinszky:jcp:2011} which is caused by an overestimation of the correlation energy at the short range. 
Binding energies, on the other hand, show a tendency to be underestimated.\cite{Furche:2001,Janesko/Henderson/Scuseria:2009,Toulouse/etal:2009,Harl/Kresse:2008,Harl/Kresse:2009,Rohlfing/Bredow:2008,Ren/Rinke/Scheffler:2009,harl:prb:10,ren:prl:11}
Moreover, the RPA correlation energy is not self-correlation free. \cite{henderson:molphys:10,paier:jcp:10,grueneis:jcp:09} 

It has been demonstrated that the overestimation of the 
absolute correlation energy can be almost entirely removed by 
adding a second-order screened exchange (SOSEX) term.\cite{freeman:prb:77,grueneis:jcp:09,henderson:molphys:10}
For one-electron systems the self-interaction error in EX+\cRPA\ 
is exactly canceled by adding this term,\cite{grueneis:jcp:09,henderson:molphys:10}
however, for systems with more than one electron, a many-electron self-interaction 
error\cite{Manyes,MCY_on_SIE} prevails. \cite{henderson:molphys:10}
%
The SOSEX can also be interpreted
as a correction to the RPA correlation energy that can
be included to {\em approximately} restore the antisymmetry of the many
electron description.\cite{henderson:molphys:10}
%
Furthermore, SOSEX 
improves binding energies, although a sizeable underestimation persists.\cite{freeman:prb:77,hu:prb:86,kresse:elba,grueneis:jcp:09,henderson:molphys:10,paier:jcp:10}
The underbinding problem can also  be alleviated, in particular for weakly 
interacting systems, by adding a correction deriving from single excitations (SE) \cite{ren:prl:11} to EX+\cRPA\
built on a reference state obtained from Kohn-Sham (KS) density-functional theory (DFT).
This suggests that \cRPA(+SOSEX) yields good estimations for the correlation energy
but errors in the exchange energy are sizeable if Kohn-Sham orbitals are used
to evaluate the exact exchange.
In light of these observations it is timely to extend the critical assessment of EX+\cRPA\ to a wider 
class of systems and to consider combinations of the corrections suggested
before. 
In this paper we will address this objective
 by performing benchmark calculations for atomization energies on an 
appreciable test set of archetypal insulating solids and small 
molecules, \cite{curtiss:jcp:97,pople:jcp:89,curtiss:jcp:89,curtiss:jcp:98} as well as reaction
 and activation energies for hydrogen and non-hydrogen transfer reactions.\cite{BH42,NHTBH38} 
The schemes we include are EX+\cRPA\ based on KS-DFT reference states, and those beyond EX+\cRPA\ by adding
corrections from SE or SOSEX individually, or both of them. In addition we will also assess the hybrid-type schemes
\cite{ren:prl:11} where one replaces the total energy at the EX level evaluated with  KS-DFT orbitals by that evaluated with 
Hartree-Fock (HF) orbitals, as an effective way to approximate the SE contribution.\cite{janesko:jcp:08}
The second-order single excitation correction can diverge when the gap between 
occupied and virtual states closes, with detrimental effects for the description of the transition states in 
chemical reactions. As briefly discussed in Ref.~\onlinecite{ren:prl:11}, including higher-order terms
in the spirit of \cRPA\ permits a resummation of the SE correction, as will be 
demonstrated in Section \ref{sec:renorm}. 
This so called renormalized SE (\cRSE) is well behaved and is included in our benchmark tests.

The paper is organized as follows: 
Sections \ref{sec:theory} and \ref{sec:compdet} briefly summarize the important aspects of the underlying theory 
and the computational parameters of our work. 
Results on molecular and solid-state atomization energies as well as reaction energies and barrier heights 
are presented in  Section \ref{sec:res} before we 
draw conclusions in Section \ref{sec:conc}.

\section{Theory\label{sec:theory}}
\subsection{Basics on RPA\label{sec:basics}}

In order to properly position the methods applied in the present work within the formal framework of DFT, we
briefly recapitulate essential equations and outline the structure of the functionals used.
Currently, for total energy calculations, RPA-based functionals usually use either KS-DFT reference states, {\it i.e.}~single-particle wavefunctions
and eigenvalues or generalized KS (GKS)\cite{GKS} reference states to compute the {\em nonlocal} EX energy as well as
the {\em nonlocal} correlation energy.\cite{Furche:2001,furche:jcp:05,Janesko/Henderson/Scuseria:2009}
Within this context, the total energy is defined as
\begin{equation}
E[n] = T_{\rm s}[\{\phi_i\}] + E_{\rm H}[n] + E_{\rm ext}[n] + E_{\rm x}[\{\phi_i\}] +  E_{\rm c}[\{\phi_i\}],
\end{equation}
%
where the terms deriving from the potential contributions in the Hamiltonian,
 $E_{\rm H}$, the electrostatic Hartree or Coulomb energy and  $E_{\rm ext}$, the (external) electron-ion interaction 
depend on the local density, whereas the last two terms,  EX energy $E_{\rm x}$ and correlation energy $E_{\rm c}$, are nonlocal contributions.
Note that the KS kinetic energy, in analogy to the exact-exchange
   energy, is not an explicit functional of the density, but rather
   of the KS orbitals.
%
The nonlocality in $E_{\rm x}$ is due to the nonlocal exchange operator acting on each (occupied) orbital $\phi_{i\sigma}(\fr)$ associated to  spin $\sigma$
and its well known dependence on the (nonlocal) reduced one-particle density matrix $\rho_\sigma(\fr,\fr') = \sum^{\rm occ}_{j} \phi_{j\sigma}(\fr)\phi^\ast_{j\sigma}(\fr')$
reads
\begin{equation}
E_{{\rm x},\sigma} = -\frac{e^2}{2}\iint \frac{|\rho_\sigma(\fr,\fr')|^2}{|\fr-\fr'|}\,d^3\fr\,d^3\fr'.
\end{equation}

In contrast to the optimized effective potential (OEP) method,\cite{sharp:pr:53,talman:pra:76,casida:pra:95} in HF theory 
the exchange operator is fully nonlocal, 
and the action of the exchange operator on a single-particle wavefunction ({\it i.e.}~orbital) depends on the value of that very
orbital throughout the entire space (see Ref.~\onlinecite{szabo}).
Note that the correlation energy $E_{\rm c}$ is a functional of both occupied as well as unoccupied eigenstates
and requires knowledge of the associated eigenenergies as well (see below).
However, both, $E_{\rm x}$ and $E_{\rm c}$ are {\it implicit} functionals of the electron density $n$
(see {\it e.g.}~Ref.~\onlinecite{DreizlerGross}).
Recent work pursuing the construction of a {\em local} RPA correlation potential are presented in 
Refs.~57$-$62.
%
Work in this direction is of great value, since it ultimately enables calculations of a self-consistent RPA correlation energies staying rigorously within the KS-DFT picture.

The RPA correlation energy can be conveniently derived from (i) perturbation theory or (ii) from
the adiabatic-connection fluctuation-dissipation (ACFD) theorem.\cite{langreth:ssc:75,gunnarsson:prb:76,langreth:prb:77}
Fundamental to the formalism is the adiabatic connection between the Hamiltonian $\hat{H}$ of an \emph{interacting} many-electron system
and the corresponding {\em noninteracting} KS Hamiltonian $\hat{H}_{\rm KS}$.
Formally, both systems may be simultaneously described by a coupling constant dependent Hamiltonian $\hat{H}(\lambda)$
with $\lambda$ being the coupling-constant or the scaling factor in the electron-electron interaction, $v_\lambda=\lambda v(\fr-\fr')$.
The electrons move in a $\lambda$-dependent external potential $v^\lambda_{\rm ext}(\fr)$.
Note that the ground-state density of $\hat{H}(\lambda)$ for all $\lambda \in [0,1]$ is constant and 
equals the physical
ground-state density $n(\fr)$, {\it i.e.}~the ground-state density of the real system.
 $\hat{H}(\lambda = 1)$ is the physical many-electron Hamiltonian with $v_{\lambda = 1}(\fr) = v_{\rm ext}(\fr)$, and $\hat{H}(\lambda = 0)$
is the KS Hamiltonian with $v_{\lambda = 0}(\fr) = v_{\rm KS}(\fr) = v_{\rm ext}(\fr) + v_{\rm H}(\fr) + v_{\rm xc}(\fr)$.
$v_{\rm H}(\fr)$ is the electrostatic Hartree potential and $v_{\rm xc}(\fr)$ is
 the xc potential.
Within ACFD, the {\it exact} KS correlation energy can be written as
\begin{equation}
\begin{split}
E_c &= -\int_0^\infty \frac{du}{2\pi}\int_0^1 d\lambda \int d{\fr} \int d{\fr'},\{ \nu(\fr-\fr') \times \\
&(
 \chi_\lambda(\fr,\fr';iu) - \chi_0(\fr,\fr';iu)             )\}\, .
\end{split}
\label{eq:acfd}
\end{equation}
Here $\nu(\fr-\fr')=1/|\fr-\fr'|$ is the bare Coulomb interaction kernel, and $\chi_0$ is the
KS independent-particle response function at imaginary frequencies $iu$,
\begin{equation}
\chi_0(\fr,\fr';iu) = 2\sum_{i}^\text{occ}\sum_{a}^\text{unocc}\frac{\phi^*_i(\fr)\phi_a(\fr)\phi^*_a(\fr')\phi_i(\fr')}{iu+\eps_i-\eps_a} + c.c.\,,
\label{eq:chi_0}
\end{equation}
where $c.c.$ denotes ``complex conjugate" and the prefactor 2 acounts for the spin-degeneracy in closed-shell systems. 
In Eq.~(\ref{eq:acfd}), $\chi_\lambda$  is the density-density response function of the
``intermediately'' interacting many-electron system employing a scaled Coulomb potential $\nu_\lambda$.
We adhere to the commonly used notation of $i,j \ldots$ being occupied, {\it i.e.}~hole KS states and
$a,b \ldots$ being unoccupied or virtual, particle states.
In principle, a Dyson-type integral equation\cite{gross:prl:kohn} has to be solved for $\chi_\lambda$,
\begin{equation}
\label{eq:dys}
\chi_\lambda = \chi_0 + \chi_0 \, (\nu_\lambda + f^\lambda_{\rm xc})\, \chi_\lambda\, ,
\end{equation}
with $f^\lambda_{\rm xc}$ as the xc kernel, {\it i.e.}~the functional derivative of the
exchange-correlation potential with respect to the density.
%
%
Within RPA, $f_{\rm xc} = 0$, {\it i.e.}~using many-body terminology,\cite{RubioReining} 
so-called vertex corrections are not included in the response function $\chi$ or equivalently in the screening of the Coulomb interaction.
Solving Eq.~(\ref{eq:dys}) for $\chi_\lambda$ with $f^\lambda_{\rm xc}=0$ corresponds to the diagrammatic resummation of ring graphs 
\cite{freeman:prb:77,harris} to infinite order.
%
In passing we note that, working within RPA, Eq.~(\ref{eq:dys}) can be rearranged to
\begin{equation}
\label{eq:geom}
\chi_\lambda = (1 - \chi_0 \nu_\lambda)^{-1}\cdot\chi_0 = [ 1 + \chi_0 \nu_\lambda + \chi_0 \nu_\lambda \chi_0 \nu_\lambda + \ldots ]\cdot \chi_0,
\end{equation}
reflecting the above mentioned summation of the (screened) Coulomb interaction up to infinite order in $\chi_0 \nu_\lambda$.
As will be seen later, Eq.~(\ref{eq:geom}) resembles the coupled-cluster amplitude equations where
so-called particle-particle, particle-hole, and hole-hole ladder terms have been removed (see Eq.~\ref{eq:ric}).
Starting from Eq.~(\ref{eq:geom}), the $\lambda$-integral is readily done and the final expression for the RPA correlation energy reads
\begin{equation}
E_c^{\rm RPA} = \int_0^\infty \frac{du}{2\pi} {\rm Tr}\{ \ln(1 - \chi_0(iu)\nu) + \chi_0(iu)\nu \}\, .
\end{equation}

\subsection{From coupled-cluster theory to RPA and RPA+SOSEX\label{sec:diagr}}
From a DFT purist's point of view, the previously outlined ACFD terminology for the RPA is certainly
the most consistent way to classify ``RPA'' as a correlation energy {\em functional} to the many-electron
ground-state.
An alternative formulation of the RPA may be motivated starting from many-body theory.
Many-body or equivalently field-theoretical diagrammatic techniques originally developed in 
quantum electrodynamics and nuclear physics\cite{brandow:rmp:67}
have been applied to the homogeneous electron gas as well as finite systems like atoms and molecules
for several decades already.
For systems that are not strongly correlated, the most successful diagrammatic, partial summation technique
(see Refs.~\onlinecite{DreizlerGross} and \onlinecite{bartlett:rmp:07}) 
is the coupled cluster (CC) expansion of the many-electron wavefunction.
The CC expansion
 to the homogeneous electron gas has been applied by Freeman,\cite{freeman:prb:77} K\"ummel, L\"uhrmann, and
Zabolitzky,\cite{kummel:prep:78} as well as Bishop and L\"uhrmann.\cite{luhrmann:prb:78,luhrmann:prb:82}
The same CC expansion techniques are indispensable ingredients for highly accurate molecular calculations.
Here, pioneers have been \v{C}\'i\v{z}ek,\cite{cizek:jcp:66,cizek:adcp:69} Paldus {\it et al.},\cite{paldus:pra:72} and Bartlett and Purvis\cite{bartlett:ijqc:78} to name a few.
A more complete list of references may be found in the recent review article by Bartlett and Musia{\l}.\cite{bartlett:rmp:07}

The CC expansion relies on the ansatz for the many-electron wavefunction, $\ket{\Psi}$,
\begin{equation}
\label{eq:cce}
\ket{\Psi} = e^{\hat{T}}\ket{\Phi},
\end{equation}
to generate the exact ground state from the ground state $\ket{\Phi}$ of the reference system
commonly within the HF approximation.
Note that $\hat{T}$ may be represented by a sum of single, double, and higher-order excitation operators, 
generating in a similar way to configuration interaction (CI)
techniques, singly, doubly substituted determinants based on the HF reference wavefunction $\ket{\Phi}$.
However, the CC expansion is distinct from CI by virtue of the exponential ansatz used in CC expansions (Eq.~\ref{eq:cce}) for the wavefunction $\ket{\Psi}$, with
\begin{equation}
\label{eq:expT}
e^{\hat{T}} = 1 + \hat{T} + \frac{1}{2!}\hat{T}^2 + \frac{1}{3!}\hat{T}^3 + \ldots,
\end{equation}
introducing so-called disconnected products of excitations responsible for the size-extensivity of
the coupled cluster correlation energy.\cite{bartlett:arpc:81}

In coupled-cluster doubles theory (CCD) the excitation operator corresponds to a double excitation
operator only, where
\begin{align}
\hat{T} &\equiv \hat{T}_2 \quad \text{with}\label{eq:t1plust2}\\
\hat{T}_2 \ket{\Phi} &= \sum_{i < j }^{ N_{\rm occ.}}\sum_{a < b}^{ N_{\rm virt.}} t_{ij}^{ab} \ket{\Phi_{ij}^{ab}}.
\end{align}
The amplitudes $t_{ij}^{ab}$ are obtained from solving a set of so-called doubles amplitude equations reading
\begin{equation}
\langle \Phi_{ij}^{ab} | e^{ -\hat{T}} \hat{H}  e^{ \hat{T}} |\Phi \rangle = 0.
\label{eq:amplitudeequationsf2}
\end{equation}
Solving Eq.~(\ref{eq:amplitudeequationsf2}) self-consistently for $t_{ij}^{ab}$
results in a resummation of infinitely many diagrams of a certain type. 
Removing all terms from the above
amplitude equation that do not correspond to so-called ring-diagrams defines the so-called ring-CCD.

Recently, the equivalence between direct, {\it i.e.}~``Coulomb term only'' 
ring-CCD (drCCD) and RPA as considered by Freeman,\cite{freeman:prb:77}
reexamined by Gr{\"u}neis and Kresse\cite{kresse:elba} and Scuseria {\it et al.},\cite{Scuseria/Henderson/Sorensen:2008}
has been demonstrated.
Scuseria {\it et al.}~algebraically showed that the CCD approximation to the many-electron wavefunction contains the ring-approximation, {\it i.e.}~the
RPA to the ground-state correlation energy, but also includes selected higher-order exchange and ladder
diagrams.\cite{freeman:prb:77,luhrmann:prb:78,luhrmann:prb:82}
In other words, RPA equals drCCD and  therefore corresponds to a subset of CCD diagrams.

Within the framework of CC expansions, the RPA and RPA+SOSEX correlation energies may be
calculated using drCCD amplitudes $\{t_{ij}^{ab}\}$ by employing the respective equations,\cite{freeman:prb:77,Scuseria/Henderson/Sorensen:2008,grueneis:jcp:09}
\begin{align}
E_c^{\rm RPA} &= \frac12 \sum_{ijab} B_{ia,jb} \,t_{ij}^{ab}\\
E_c^{\rm RPA+SOSEX} &= \frac12 \sum_{ijab} K_{ia,jb}\, t_{ij}^{ab}.
\end{align}
The matrices $B_{ia,jb}$ and $K_{ia,jb}$ are of rank $N_{\rm occ}\times N_{\rm virt}$, and they are defined
by two-electron integrals $B_{ia,jb} = \braket{ij}{ab}$ and $K_{ia,jb} = \braket{ij}{ab} - \braket{ij}{ba}$,
respectively,
\begin{equation}
\label{eq:twoint}
\langle pq | rs \rangle = \iint \phi_p^\ast(\fx)\phi_r(\fx) \frac{1}{|\fr - \fr'|}\phi_q^\ast(\fx')\phi_s(\fx')d\fx d\fx',
\end{equation}
\noindent
with $\fx$=$\{\fr,\sigma\}$.
The amplitudes $\{t_{ij}^{ab}\}$ are obtained from solving a set of nonlinear Riccati equations,
closely related to
 to the time-dependent HF or more precisely the time-dependent Hartree method,\cite{Scuseria/Henderson/Sorensen:2008}
\begin{equation}
\label{eq:ricc}
\begin{split}
\langle ij | ab \rangle &+ (\eps_c - \eps_k) \delta_{ac} \delta_{ik} t_{kj}^{cb} + \langle ic | ak \rangle t_{kj}^{cb}\\
                        &+   t_{ik}^{ac} (\eps_c - \eps_k) \delta_{bc} \delta_{jk} + t_{kj}^{cb} \langle ic | ak \rangle \\
                        &+ t_{ik}^{ac} \langle kl | cd \rangle t_{lj}^{db} = 0\, .
\end{split}
\end{equation}
The previous equation can be rewritten in a more compact form,\cite{Scuseria/Henderson/Sorensen:2008}
\begin{equation}
\label{eq:ric}
\B + \A \T + \T \A + \T \B \T = 0,
\end{equation}
with $A_{ia,jb}=(\epsilon_a - \epsilon_i)\delta_{ij}\delta_{ab} + \langle ib|aj \rangle$, $B_{ia,jb}=\langle ij|ab \rangle$, and $T_{ia,jb}=t_{ij}^{ab}$,
underlining the quadratic order in the amplitudes' matrix ${\bf T}$.

Freeman has evaluated the RPA correlation energy of the unpolarized electron gas for various electron
densities\cite{freeman:prb:77} using the drCCD equations
and compared them to Hedin's RPA results (see Table~II in Ref.~\onlinecite{hedin:pr:65})
following an approach suggested by Nozi{\`e}res and Pines.\cite{nozieres:nuocim:58}
%

%
Both agree to within the numerical accuracy employed in the calculations.
Moreover, Freeman has gone beyond RPA via inclusion of the second-order screened exchange (SOSEX) diagram.
He found that SOSEX reduces the correlation 
energy by about 30\%.
Monkhorst and Oddershede came to similar conclusions employing RPA and RPA+SOSEX to metallic hydrogen,\cite{monkhorst:prl:73} and Gr{\"u}neis observed a similar reduction of the correlation energy for small atoms\cite{grueneis:jcp:09}
finding good agreement with highly accurate coupled cluster correlation energies only after inclusion of
SOSEX. 
Finally we note that until recently the formulation of SOSEX within an ACFD framework has not been entirely
clear, but has lately been shown by Jansen {\it et al.}.\cite{jansen:jcp:10}

%
%
%
%

\begin{table*}[htb]
\renewcommand{\arraystretch}{1.2}
\setlength{\tabcolsep}{3pt}
\parbox{.95\textwidth}{
\caption{\label{tab:acr}
The list of methods used throughout this work and their acronyms.
Note that the total energy at the exact-exchange level is abbreviated by ``EX''.
}
}
\begin{tabularx}{0.95\textwidth}{@{\hspace{3mm}}l@{\hspace{3mm}}X}
\hline
\hline
(EX+\cRPA)@PBE           & EX and \cRPA\ evaluated with a PBE reference, {\it i.e.}~PBE orbitals and eigenvalues\\
HF+\cRPA @PBE            & HF total energy combined with \cRPA\ using a PBE reference \\
(EX+\cRPA+\cSE)@PBE        & EX and \cRPA\ augmented with \cSE\ using the PBE reference  \\
(EX+\cRPA+\cRSE)@PBE       & EX and \cRPA\ augmented with \cRSE\ using the PBE reference    \\
HF+(\cSOSEX)@PBE     & HF total energy combined with  \cSOSEX\ using the PBE reference \\
(EX+\cSOSEX)@PBE     & EX, \cSOSEX\ using the PBE reference \\
(EX+\cSOSEX+\cRSE)@PBE & EX, \cSOSEX, and \cRSE\ using the PBE reference \\
\hline
\hline
\end{tabularx}
\end{table*}

\subsection{Single excitations and their renormalization\label{sec:renorm}}
As alluded to above, in most practical calculations, RPA and SOSEX correlation energies are 
evaluated using KS orbitals from local or semilocal density functionals, \cite{Furche:2001,Harl/Kresse:2008} or generalized
KS orbitals \cite{Ren/Rinke/Scheffler:2009,Janesko/Henderson/Scuseria:2009,paier:jcp:10} from range-separated density functionals. This way, both RPA and SOSEX can be interpreted
in terms of many-body perturbation theory (MBPT) based on a (generalized) KS reference state, where only a 
selected type of diagrams are summed up to infinite order. If one performs a simple Rayleigh-Schr{\"o}dinger 
perturbation theory (RSPT) starting from an (approximate) KS-DFT reference, and examines the perturbation series
at second order, one can identify a term arising from single excitations (SE), that is not included in RPA or
SOSEX correlation energies. In terms of single-particle orbitals, this term can be expressed as
 \begin{equation}
   E_c^\text{SE} = \sum_{ia} \frac{|\langle i|v^\text{HF}- v^\text{eff}|a\rangle|^2}{\epsilon_i - \epsilon_a},
   \label{eq:E_c_SE}
 \end{equation}
where $v^\text{HF}$ is the self-consistent HF potential, and $v^\text{eff}$ is the effective 
single-particle potential that defines the non-interacting reference Hamiltonian $h^\text{eff}$ giving rise to the
single-particle orbitals $|i\rangle$ and $|a\rangle$ in the above expression. (See the supplemental material of
Ref.~\onlinecite{ren:prl:11} for a detailed derivation.) As is obvious from Eq.~(\ref{eq:E_c_SE}),
$E_c^\text{SE}$ trivially vanishes for the HF reference, {\it i.e.}, when $v^\text{eff}=v^\text{HF}$, but is nonzero
otherwise. It has been shown that adding this term to RPA improves the description of weak interactions
significantly.\cite{ren:prl:11} Note that the choice of $v^\text{eff}$ in Eq.~(\ref{eq:E_c_SE}) is slightly 
different in RSPT from that in the 2nd-order G{\"o}rling-Levy perturbation theory (GL2). \cite{gorlinglevy:prb:93} 
In the latter case, $v^\text{eff}=v^\text{EXX-OEP}$, with $v^\text{EXX-OEP}$ being the exact-exchange OEP 
\cite{sharp:pr:53,talman:pra:76,casida:pra:95} potential. The difference of the two perturbation theories lies in 
the choice of the adiabatic-connection path ($\lambda$-integral) -- in GL2 the electron density is kept fixed  
along the path way and the perturbative Hamiltonian has a non-linear dependence on $\lambda$, whereas in RSPT the 
$\lambda$-dependence of the perturbative Hamiltonian is linear, but the electron density varies along the $\lambda$-integral.
Eq.~(\ref{eq:E_c_SE}) in RSPT is more efficient and practically useful in the sense that there is no need to solve 
the computationally intensive and sometimes numerically problematic EXX-OEP equation and more flexible 
in the sense that it can be matched to any suitable reference state. The price one has to pay is that the theory, 
strictly speaking, is not KS-DFT formulated within the ACFD framework.

The SE contribution at second order as given by Eq.~(\ref{eq:E_c_SE}) may become ill-behaved when the single-particle
gap closes. To deal with this problem, in Ref.~\onlinecite{ren:prl:11} a sequence 
of higher-order terms involving SE processes have been identified and summed up in the spirit of RPA. 
This leads to a ``renormalized" SE (\cRSE) contribution to the correlation energy, 
 \begin{equation}
   E_c^\text{RSE} = \sum_{ia} \frac{|\langle i|\Delta v|a\rangle|^2}
        {\epsilon_i - \epsilon_a + \langle i| \Delta v |i\rangle - \langle a | \Delta v | a \rangle },
   \label{eq:E_c_RSE}
 \end{equation}
where $\Delta v = v^\text{HF} - v^\text{eff}$. The additional term  $\langle i| \Delta v |i\rangle - \langle a | \Delta v | a \rangle$ in the denominator of Eq.~(\ref{eq:E_c_RSE}) is negative definite, and prevents the
possible divergence of the expression even when the KS gap closes. The \cRSE\ correction is therefore expected to have a 
more general applicability, while preserving the good performance of the 2nd-order SE for wide-gap molecules and insulators. In deriving Eq.~(\ref{eq:E_c_RSE}), however, the ``non-diagonal" elements in the higher-order SE diagrams have been 
neglected for simplicity. Such an approach lacks invariance with respect 
to unitary transformations (orbital rotations) within the occupied and/or unoccupied subspaces. The 
orbital-rotation-invariance can be restored by including also the ``non-diagonal" elements.
This can be achieved by first semi-diagonalizing the Fock Hamiltonian $f=h^\text{eff}+v^\text{HF}-v^\text{eff}$
separately within the occupied and unoccupied subspaces of $h^\text{eff}$ and utilizing the resultant  
(so-called \emph{semi-canonical}) orbitals and orbital energies in Eq.~(\ref{eq:E_c_SE}). A detailed
description of this procedure will be presented in a forthcoming paper. 
However, we emphasize that results presented in this work are based on Eq.~(\ref{eq:E_c_RSE}), but despite the lack
of rotational invariance in the orbitals of this approach, numerical results are only very little affected.

As also demonstrated in Ref.~\onlinecite{ren:prl:11}, the SE contributions to the correlation energy
can be effectively accounted for to a large extent by replacing the non-self-consistent HF total energy computed
using KS orbitals by its self-consistent counterpart. In this so-called hybrid-RPA scheme 
the RPA correlation energy is still evaluated using KS orbitals, whereas the EX term
is evaluated using HF orbitals. The same strategy can be applied to  ``RPA+SOSEX'' calculations. In this work, we
will benchmark the influence of SE contributions on the performance of RPA and SOSEX both by explicitly 
including the (r)SE corrections and in terms of the hybrid scheme.

 As outlined in Ref.~\onlinecite{ren:prl:11} by Ren {\it et al.}, rendering the energy functional
stationary with respect to variations in the orbitals implies a zero correlation energy contribution
stemming from SEs.
This is well known as Brillouin's theorem.
It will be demonstrated in this work, that SE effects represent a non-negligible contribution to the
correlation energy and consequently affect results on thermochemistry and kinetics.
In the field of quantum chemistry effects induced by SEs are known as orbital-relaxation effects.
\cite{raffenetti:tca:92,ruedenberg:ijqc:76}
Besides MBPT discussed above, the SE terms are present in the CC theory as well.
In this context, Scuseria and Schaefer have shown that CCD employing 
optimized-orbitals
(see Ref.~\onlinecite{scuseria:cpl:87}) gives results very close to CCSD.
On the other hand, optimizing orbitals for  CCSD calculations does not lead to significant improvements in the wavefunction.
In other words, changes in the correlation energy induced upon inclusion of SEs may be effectively incorporated
by means of a unitary transformation, {\it i.e.}~rotation of the orbitals, as given in Eq.~(6) of Ref.~\onlinecite{scuseria:cpl:87}.

We close this section by presenting Table~\ref{tab:acr}, which summarizes the acronyms of the various methods applied 
in this work.
For the KS single-determinant reference wave function we use the Perdew, Burke, and Ernzer\-hof (PBE)\cite{PBE} generalized gradient approximation (GGA).
We adopt the notation introduced by Ren {\it et al.} in Ref.~\onlinecite{Ren/Rinke/Scheffler:2009}, hence
``@PBE'' means ``evaluated using PBE orbitals and orbital energies''.
%
This particular choice of orbitals is mainly driven by the following arguments: (i) 
PBE contains no empirically adjusted parameters, (ii) performs 
slightly better than LDA (see {\it e.g.}~Ref.~\onlinecite{Harl/Kresse:2008}), and (iii) it is computationally less expensive to calculate the orbitals using semilocal functionals
instead of {\it e.g.}~hybrid functionals.\cite{Ren/Rinke/Scheffler:2009}
In addition, once one restricts the input orbitals to KS orbitals,
results have shown to be virtually identical to those obtained using PBE orbitals.\cite{eshuis:jpcl:2011,eshuis/bates/furche:tca:12}

\section{Computational Details\label{sec:compdet}}

Computational results of the present work are based on calculations using
(i) the Vienna {\it ab initio} simulation package 
\VASP,\cite{kresse:prb:93,kresse:96,kresse:96_2} (ii)
a development version
of the \GDV\cite{gdv-g1} suite of programs, and (iii) \AIMS. \cite{blum:cpc:09,ren:prb:sub}   
All of the software packages used have the RPA and RPA+SOSEX functionals available since 
recently.\cite{Harl/Kresse:2008,Janesko/Henderson/Scuseria:2009,paier:jcp:10,Ren/Rinke/Scheffler:2009}
\VASP\ uses periodic boundary conditions and projector augmented plane waves as 
a basis set, which makes it ideally suited for extended, crystalline systems.
\GDV\ is based on local, analytic Gaussian type (GT) basis functions
using open boundary conditions and the linear combination of atomic orbitals to expand the molecular orbitals.
\AIMS\ primarily uses numeric, atom-centered basis functions, but  GT orbitals can be employed as well. 
In both cases, all the required integrals are evaluated numerically on an overlapping atom-centered grid.\cite{blum:cpc:09} The resolution-of-identity approximation is used to handle the four-centered Coulomb repulsion integrals and the KS response function (details of the implementation have been presented in Ref. \onlinecite{ren:prb:sub}). 
In this work GT orbitals are used in \AIMS\ calculations to facilitate a direct comparison with \GDV\ and 
the extrapolations to the complete basis set (CBS) limit. 
%

%
In this work we present statistical errors for the G2-1 set,\cite{curtiss:jcp:97,pople:jcp:89,curtiss:jcp:89,curtiss:jcp:98} as well as for BH6,\cite{ae6bh6} HTBH38/04, and NHTBH38/04 sets of 38 hydrogen transfer and
38 nonhydrogen transfer barrier heights after Zhao {\it et al.}\cite{BH42,NHTBH38}
Results for the molecular test sets use a  two-point extrapolation procedure
on the correlation energies to attain the complete basis set (CBS) 
limit.\cite{kutzelnigg:jcp:92,helgaker:jcp:97,halkier:cpl:98} 
The chosen ansatz is
motivated by an atomic partial wave expansion of the two-particle many-body wavefunction, \cite{helgaker:jcp:97}
\begin{equation}
\label{eq:ccsd_ext}
E_{\rm corr}^{X} =  E^{\infty}_{\rm corr} + \frac{a}{X^3},
\end{equation}
where the $ E_{\rm corr}^{X}$ are correlation energies corresponding to the  cc-pVXZ basis sets.
For G2-1, CBS calculations are based on Dunning's correlation-consistent 
cc-pVQZ and cc-pV5Z basis sets.\cite{dunning:jcp:89,woon:jcp:93}
Note that throughout this work CBS extrapolation will be denoted by, e.g., cc-pV(Q,5)Z.

Moreover, G2-1 calculations employ the Boys-Bernardi counterpoise correction\cite{boys:molphys:70} 
to correct for  basis set superposition errors (BSSE) within a particular basis set.
Therefore, we emphasize that the CBS procedure uses BSSE free correlation energies.
In order to avoid inaccuracies  from numerical  quadrature of xc energy contributions,
\GDV\ calculations use a grid of 400 radial shells and 770 angular points in each shell to converge the KS orbitals.
\GDV\ employs a root-mean-square
convergence criterion for the density matrix in the SCF
iteration of 0.1 $\mu$Hartree, which
implies an energy convergence no worse than at least
0.01 $\mu$Hartree\,(\GDV\ keyword: \verb+SCF=tight+).
In \AIMS\ the  grid setting ``\texttt{tight}" together with ``\texttt{radial\_multiplier=6}" 
has been used  to achieve convergence within one $\mu$Hartree.

\begin{table}[htb]
\caption{
Matching radii $r_c$ of the PAW potentials used in the
present work. If the matching radii differ
for specific quantum numbers, they are specified for each $l$-quantum number
using subscripts.
}
\label{tab:pseudos}
\begin{ruledtabular}
\begin{tabular}{lcc|lcc}
& Valence & $r_c$ [a.u.]         &  & Valence & $r_c$ [a.u.] \\ \hline
H  & 1s   & 1.0$_{s}$ 1.1$_{pd}$ & F  & 2s2p  & 1.1$_{s}$ 1.4$_{pd}$ \\
Li & 1s2s & 1.2$_{s}$ 1.5$_{pd}$ & Mg & 2p3s  & 2.0$_{sd}$ 1.6$_{p}$   \\
B  & 2s2p & 1.5$_{s}$ 1.7$_{pd}$ & Al & 3s3p  & 1.9$_{spd}$ 2.0$_{f}$\\
C  & 2s2p & 1.2$_{s}$ 1.5$_{pd}$ & Si & 3s3p  & 1.5$_{s}$ 1.9$_{pd}$  \\
N  & 2s2p & 1.3$_{s}$ 1.5$_{pd}$ & P  & 3s3p  & 1.9$_{sp}$ 2.0$_{df}$\\
O  & 2s2p & 1.2$_{s}$ 1.5$_{pd}$ & Cl & 3s3p  & 1.7$_{s}$ 1.9$_{pdf}$  \\
\end{tabular}
\end{ruledtabular} 
\end{table}

Results on barrier heights in BH6, HTBH38/04, and NHTBH38/04 use a cc-pV(T,Q)Z CBS extrapolation of
the correlation energies and do not employ counterpoise corrections. 
To test for the errors incurred,
we compare with  benchmark results obtained using RPA and RPA+SOSEX given in Ref.~\onlinecite{paier:jcp:10}.
The statistical errors in barrier heights deviate from the aforementioned benchmark values
 by at most 1 kJ/mol.
Hence, the errors incurred using smaller basis sets are minute and consequently are not expected to bias
the conclusions.

\begin{table}[t] \caption{\label{tab:lattice_constants} Experimental lattice constants, $a^{\rm exp}$, extrapolated to 0~K. 
Energy cutoffs for the one-electron wave functions $E_{\rm PW}$ as well as energy cutoffs for representing the overlap charge 
densities $E_{\chi}$ employed in the calculation of the atomization energies of solids. The corresponding structures are denoted 
using the Strukturbericht symbols in parenthesis in the first column (A4=diamond, B1=rock-salt, B3=zinc-blende). 
All energies and lattice constants in eV and \AA, respectively.} 
\begin{ruledtabular} 
\begin{tabular}{lccc} 
& $a^{\rm exp}$ &$E_{\rm PW}$ & $E_{\chi}$ \\ 
\hline 
C (A4) & 3.567$^{\rm a}$ & 550 & 400 \\ 
Si (A4) & 5.430$^{\rm a}$ & 450 & 300\\ 
SiC (B3) & 4.358$^{\rm a}$ & 550 & 400 \\ 
BN (B3) & 3.607$^{\rm b}$ & 550 & 400 \\ 
BP (B3) & 4.538$^{\rm b}$ & 450 & 350 \\ 
AlN (B3) & 4.380$^{\rm c}$ & 550 & 400 \\ 
AlP (B3) & 5.460$^{\rm b}$ & 450 & 350 \\ 
LiH (B1) & 4.064$^{\rm d}$ & 600 & 450 \\ 
LiF (B1) & 4.010$^{\rm a}$ & 600 & 450 \\ 
LiCl (B1) & 5.106$^{\rm a}$ & 600 & 450 \\ 
MgO (B1) & 4.207$^{\rm a}$ & 600 & 450 \\ 
\end{tabular} 
\begin{tablenotes} 
\item $^{\rm a}$Ref.~\onlinecite{staroverov:prb:04}, $^{\rm b}$Ref.~\onlinecite{madelung:sdh:04}, $^{\rm c}$ Ref.~\onlinecite{trampert:book:1998}, $^{\rm d}$ Ref.~\onlinecite{smith:japplc:68}. 
\end{tablenotes} 
\end{ruledtabular} 
\end{table} 

The test set on atomization energies for crystalline solids includes
11 archetypal semiconductors and insulators.
Specifically it comprises  C, Si, SiC, BN, BP, AlN,
AlP, LiH, LiF, LiCl, and MgO.
The projector augmented wave (PAW) pseudopotentials (technical details in Tab.~\ref{tab:pseudos}) and kinetic
energy cutoffs employed in the present
calculations 
are identical to the ones used in Ref.~\onlinecite{grueneis:jcp:10}.
Table~\ref{tab:lattice_constants} summarizes the lattice constants used in 
``post-RPA'' calculations.
Moreover, we specify plane wave cutoffs for  the
overlap charge densities described in Refs.~\onlinecite{harl:prb:10} and \onlinecite{grueneis:jcp:10}.
The SOSEX correlation energy was calculated using a ($3\times3\times3$) $\Gamma$-centered $k$ mesh,
except for BN and BP due to a slower $k$-point convergence of the energy. 
For these systems a ($4\times4\times4$) mesh was used.
RPA correlation energies are taken from the literature (see Ref.~\onlinecite{harl:prb:10}).
In \VASP, atoms are calculated using a supercell approach.
The dimension of the supercells has been chosen as ($9\times9\times9$) \AA$^3$ in size.
To reduce the computational cost of the ``RPA+SOSEX'' calculations for isolated atoms,
natural orbitals obtained using second order perturbation theory have been employed.
As outlined in Ref.~\onlinecite{gruneis:jctc:11}, natural orbitals substantially improve convergence of the correlation energy with respect to the number of virtual orbitals.

%
\begin{table}[htb]
\renewcommand{\arraystretch}{1.2}
\setlength{\tabcolsep}{8pt}
\parbox{.95\columnwidth}{
\caption{\label{tab:bench}
Benchmark calculations for atomic He using \AIMS\ and \GDV\ and  a cc-pV5Z GT orbital basis set. Results are given in Hartree atomic units.
}}
\begin{tabular}{lcccc}
\hline
\hline
 He / cc-pV5Z   &     \GDV\      &    \AIMS\   \\
\hline
HF                &  -2.86162468     & -2.86162483           \\ 
MP2              &  -0.03640606     & -0.03640651           \\ 
\cRPA@HF           &  -0.06524488     & -0.06524570           \\ 
(\cSOSEX)@HF  &   -0.03262244    & -0.03262285           \\ 
\hline
\hline
\end{tabular}
\end{table}

To assess the codes used in this work, we
compare numerical results obtained using the
``RPA'' and ``RPA+SOSEX'' implementations of \GDV\ and \AIMS.
Table~\ref{tab:bench} shows   correlation energies for the He atom obtained using the cc-pV5Z basis set.
In order to avoid errors caused by numerical integration, we
decided to use (restricted, {\it i.e.}~spin-unpolarized) HF orbitals and eigenvalues for the calculation of \cRPA\ and \cSOSEX.
The agreement found is close to perfect. Differences between results  are within a sub-micro-Hartree error margin.
In passing we mention that \AIMS\ employs the resolution-of-identity (RI) technique,\cite{ren:prb:sub} which (i) reduces the computational workload significantly and (ii), as shown in Tab.~\ref{tab:bench}, does not sacrifice accuracy.
For the molecular test sets, we always crosscheck the ``RPA" and ``RPA+SOSEX" results obtained
with the \GDV\ suite of program and \AIMS\ to make sure that the results presented in this work are
not affected by the actual implementations.
``SE" and ``\cRSE" have so far only been implemented in \AIMS\ and we use these results throughout.

\section{Results and Discussion\label{sec:res}}
\begin{table*}[htb]
\renewcommand{\arraystretch}{1.1}
\setlength{\tabcolsep}{2pt}
\newlength{\tabs}
\setlength{\tabs}{8mm}
\parbox{17.8cm}{
\caption{\label{tab:res}
Mean  errors (ME) and mean unsigned errors (MUE) in 
atomization or binding energies of 11 solids (see Tab.~\ref{tab:lattice_constants}) and 55 molecules
(G2-1), in the barrier heights comprised in BH6, in HTBH38/04 (hydrogen transfer barriers), as well as in NHTBH38/04 (non-hydrogen transfer barriers).
Results are given in kJ/mol.
}}
\begin{threeparttable}
\begin{tabular}{lrr@{\hspace{\tabs}}rr@{\hspace{\tabs}}rr@{\hspace{\tabs}}rr@{\hspace{\tabs}}rr}
\hline
\hline
  Method           &\multicolumn{2}{c}{Solids} &\multicolumn{2}{l}{\phantom{xyza}G2-1} & \multicolumn{2}{l}{\phantom{xyza}BH6} & \multicolumn{2}{l}{HTBH38} & \multicolumn{2}{c}{NHTBH38} \\
       &         ME   &   MUE  &    ME  &   MUE    &  ME   & MUE  &  ME  & MUE & ME & MUE     \\
\hline
(EX+\cRPA)@PBE          &$-$67.5 & 67.5 & $-$42.7   & 42.8$^{\rm a}$                     & \phantom{$-$}1.2   & 7.5$^{\rm a}$  &$-$0.8    & 7.1  &$-$10.5   &  12.1 \\
HF+\cRPA @PBE           &$-$34.7 & 36.7 & $-$25.3   &\phantom{1}30.3$^{\rm \phantom{a}}$ & $-$25.5            &25.5$^{\rm \phantom{a}}$  &$-$36.8   &36.8  &$-$48.5   &48.5\\
(EX+\cRPA+\cSE)@PBE     &        &      & $-$14.2   &\phantom{1}22.9$^{\rm \phantom{a}}$ & $-$23.8            &23.8$^{\rm \phantom{a}}$  & $-$52.7  & 52.7 &  $-$50.6 &  51.9 \\
(EX+\cRPA+\cRSE)@PBE    &        &      & $-$26.2   & 27.4$^{\rm \phantom{a}}$           & $-$14.8            & 16.3$^{\rm \phantom{a}}$ &$-$18.0   & 21.7 & $-$31.4  & 31.4  \\
(EX+\cSOSEX)@PBE        &$-$27.0 & 27.0 & $-$20.3   &\phantom{1}23.0$^{\rm {a}}$         &\phantom{$-$}17.6   &17.6$^{\rm a}$  & 22.2     &22.2  &13.4      &15.5 \\
HF+(\cSOSEX)@PBE        &5.8     & 17.4 & $-$2.9    &\phantom{1}13.0$^{\rm \phantom{a}}$ &$-$9.2              &9.2$^{\rm \phantom{a}}$   &$-$13.8   &13.8  & $-$24.7  &25.5  \\
(EX+\cSOSEX+\cRSE)@PBE  &        &      & $-$4.0    & 13.9$^{\rm \phantom{a}}$           & 3.1                & 3.7$^{\rm \phantom{a}}$  & 3.6      & 5.4  &  $-$6.3  & 17.6   \\
\hline
\hline
\end{tabular}
\begin{tablenotes}
\item[a] See Ref.~\onlinecite{paier:jcp:10}. Note that differences in the MUE of G2-1 are due to the different values for the experimental
dissociation energies (see Ref.~\onlinecite{feller:jcp:99}).
\end{tablenotes}
\end{threeparttable}
\end{table*}

Central findings of this work are summarized in Tab.~\ref{tab:res} presenting  binding energies in molecules (G2-1) and solids,
HT activation energies or barrier heights (BH6, HTBH38) as well as NHT barrier heights (NHTBH38).
Whenever results are compared to experiment or to the best theoretical estimates, we use mean error (ME) and mean unsigned error (MUE) as  statistical measures
 to assess the accuracy of  individual methods employed.
Note that the experimental reference values are corrected for zero point effects and are taken from
the literature (G2-1: Ref.~\onlinecite{feller:jcp:99}; atomization energies in solids: 
Ref.~\onlinecite{schimka:jcp:11}; barrier heights in BH6, HT/NHTBH38: 
Refs.~\onlinecite{ae6bh6}, \onlinecite{BH42}, and \onlinecite{NHTBH38}).
Reaction energies, as presented in Table~\ref{tab:reac_ene}, are calculated from  barrier heights in HTBH38 and NHTBH38, respectively.
\subsection{Atomization energies of small molecules and solids\label{sec:atom}}
\begin{figure}
\includegraphics[width=.99\columnwidth,clip=true]{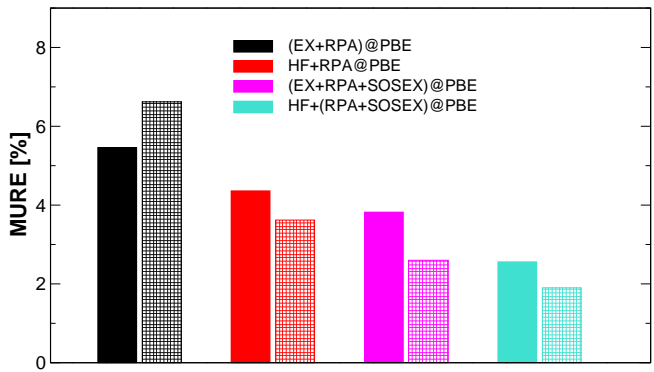}
\caption{\label{fig:molssols} 
Mean unsigned relative errors (MURE) in the atomization energies of 55 small molecules contained
in G2-1 (full bars) and 11 insulating solids (squared bars) obtained using four of the RPA-based methods presented in this
work.
}
\end{figure}

\begin{figure}
\includegraphics[width=.99\columnwidth,clip=true]{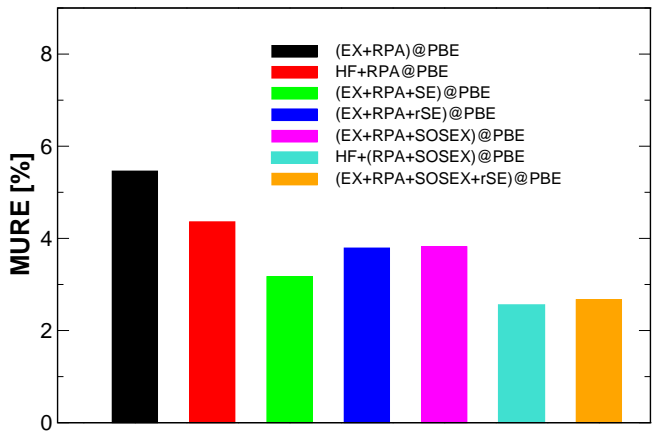}
\caption{\label{fig:g2-1} 
Mean unsigned relative errors (MURE, given in \%) in G2-1 for all RPA-based methods presented in this work.
Atomization energies use counterpoise correction and correlation energies are CBS extrapolated
using cc-pV(Q,5)Z.}
\end{figure}

The notorious underbinding of \RPA\ in molecules and solids has already been demonstrated in many studies.\cite{Furche:2001,Ren/Rinke/Scheffler:2009,Harl/Kresse:2008,Harl/Kresse:2009,grueneis:jcp:09,paier:jcp:10,harl:prb:10}
Table~\ref{tab:res} presents MEs and MUEs
in binding (atomization) energies obtained using \RPA\ for insulating solids (see Sec.~\ref{sec:compdet}) as well as for the molecules contained in the G2-1 set.
On average, \RPA\ underbinds solids compared to experiment by \unit[$-67$]{kJ/mol} (see also Ref.~\onlinecite{harl:prb:10})
and molecules by \unit[$-43$]{kJ/mol}.
We repeat that experimental binding energies are corrected for zero-point effects and are taken from the literature
(for G2-1, see Ref.~\onlinecite{feller:jcp:99}, for the test set on solids see 
Ref.~\onlinecite{schimka:jcp:11} and  references therein).

Following the suggestion of Ren {\it et al.},\cite{ren:prl:11} effects incurred by replacing the EX@PBE reference
energy by the  HF total energy have been  checked  for both
molecules and solids.
Indeed, \RPAh\ improves binding energies of molecules {\it and} solids by almost 50\% compared to \RPA.
Fig.~\ref{fig:molssols} presents  mean unsigned {\it relative} errors (MURE) in molecular (full bars) as well as solid state (squared bars) binding energies.
Overall differences in MUREs are rather small.
For the commonly applied \RPA\ method, the MURE is found to be  approximately 6\%.
Using  HF at the exact exchange level reduces the MURE by more than 1\%.
It appears that the aforementioned improvements are less pronounced at the relative scale,
and the error reduction is apparently bigger for solids than for molecules.

The explicit inclusion of the SE contribution to the correlation energy ``\cSE@PBE'' obtained using 
Eq.~(\ref{eq:E_c_SE}) has been evaluated for molecules only.
Adding ``\cSE@PBE'' to \RPA\ results in an ME of approximately \unit[$-14$]{kJ/mol} (see Tab.~\ref{tab:res}) and an MUE of approximately \unit[$23$]{kJ/mol}, clearly  outperforming \RPAh.
Relative unsigned errors in G2-1 collected in Fig.~\ref{fig:g2-1} further corroborate the improvements of \RPASE\ over \RPAh.
Overall, these results confirm the findings presented by Ren {\it et al.}~in Ref.~\onlinecite{ren:prl:11}.
However, 
``Renormalization'' of the SE contributions, as required for systems with a small one-electron
band gap in PBE (see activation energies discussed in Sec. \ref{sec:disc.barriers})
brings the atomization energies in the G2-1 test set back into almost perfect agreement with HF+RPA@PBE.
Therefore, the good agreement with experiment for the G2-1 test set on the level of  \RPASE\ is most likely
to some extent fortuitous.

As extensively discussed in Refs.~\onlinecite{grueneis:jcp:09} and \onlinecite{paier:jcp:10}, the (RPA+SOSEX) correlation energy, here denoted as ``(\cSOSEX)@PBE,''
represents a correction to \RPA\ rectifying the one-electron self-interaction error contained in ``\cRPA@PBE'' due to  exclusion principle violating diagrams.\cite{henderson:molphys:10}
Results for G2-1 obtained using \SOSEX\ are taken from Ref.~\onlinecite{paier:jcp:10} and included in Tab.~\ref{tab:res}.
The ME in G2-1 obtained using \SOSEX\ 
is approximately equal to \unit[$-20$]{kJ/mol}. For solids, the ME error reduces to \unit[$-27$]{kJ/mol}.
Compared to \RPA, this represents substantial improvements of approximately 50\% for atomization energies.

Given that both SOSEX and \cRSE, or alternatively replacing EX@PBE by HF, alleviate the tendency of \RPA\ to underbind, 
both  schemes are
expected to work cooperatively for the atomization energies of small molecules.
Indeed, replacing ``EX@PBE'' in \SOSEX\ by the HF total energy yields 
excellent results, with a slight underbinding for molecules  (ME = \unit[$-2.9$]{kJ/mol}),
and a slight overbinding for solids  (ME=\unit[$5.8$]{kJ/mol}).
Again, the HF+(\cSOSEX)@PBE (ME = \unit[$-2.9$]{kJ/mol}) and the \SOSEXRSE\ methods (ME = \unit[$-4.0$]{kJ/mol}) perform almost
on par for molecules.

In summary, single excitation diagrams improve \RPA\ atomization energies of small molecules
at virtually zero additional computational cost.
However, as we will see below, this method fails when the one-electron band gaps in PBE become small.
The better founded  rSE does not perform equally well for atomization energies when comined with RPA.
Combined with RPA+SOSEX it yields impressive atomization energies that are also in almost perfect
agreement with the ``hybrid variants''  {\it e.g.}~the (self-consistent) 
HF total energy together with ``(\cSOSEX)@PBE''. Overall this indicates that
(EX+\cSOSEX+\cRSE)@PBE or HF+(\cSOSEX)@PBE are the methods of choice for atomization energies.

\subsection{Activation energies in HTBH38 and NHTBH38 chemical reactions\label{sec:disc.barriers}}
\begin{figure}[htb]
\includegraphics[width=.85\columnwidth,clip=true]{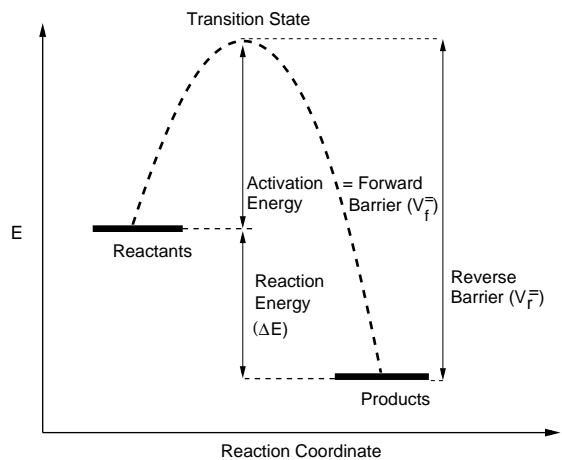}
\caption{\label{fig:activation} 
Schematic of activation and reaction energies.}
\end{figure}
The ability to accurately describe the topology of multidimensional potential energy surfaces spanned
by the internal molecular degrees of freedom, {\it i.e.}~the reaction coordinates,  in the course of a chemical reaction, is central to first principles electronic structure methods.
Calculating the  energy difference between reactants and transition states (see Fig.~\ref{fig:activation}) is
 a stringent test for the accuracy of density functionals.
As mentioned in Sec.~\ref{sec:compdet}, the HTBH38 and NHTBH38
test sets established by Truhlar and coworkers \cite{BH42,NHTBH38} will be used here to test the RPA-based functionals 
considered.

\begin{figure}[htb]
\includegraphics[width=.99\columnwidth,clip=true]{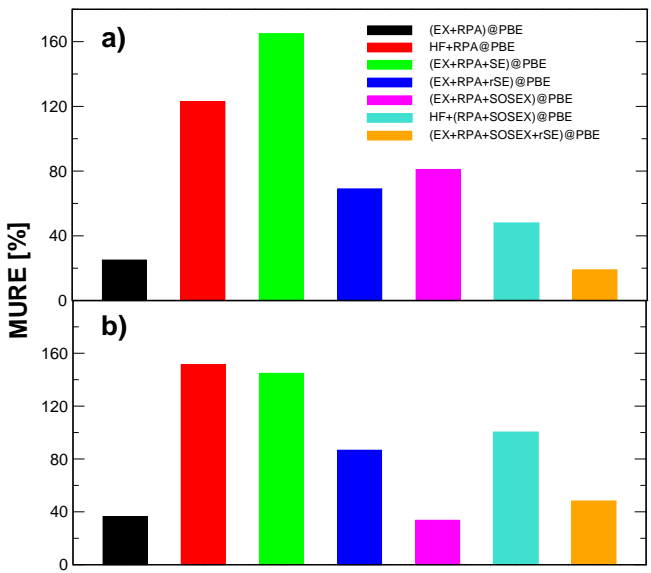}
\caption{\label{fig:barriers}
Panel a) shows mean unsigned relative errors (MURE) in HT barrier heights of HTBH38.
Panel b) shows MUREs in  NHTBH38 for the RPA-based methods presented in this work. 
Energies use a cc-pV(T,Q)Z extrapolation and
the frozen core approximation in calculated correlation energies.}
\end{figure}

Our findings on barrier heights, {\it i.e.}~activation energies (Fig.~\ref{fig:activation}), are summarized in Table~\ref{tab:res}.
MEs and MUEs
are calculated with respect to the best theoretical 
estimates currently available for HT and NHT barrier heights given in Refs.~\onlinecite{BH42}
and \onlinecite{NHTBH38}, respectively.
Furthermore the MUREs in HT barriers [panel (a)] and
NHT barriers [panel (b)] are depicted in Fig.~\ref{fig:barriers}.
Note that legends given in  Fig.~\ref{fig:barriers}
follow the color code  used in Fig.~\ref{fig:g2-1}.
To establish a  connection to
 Ref.~\onlinecite{paier:jcp:10}, Tab.~\ref{tab:res} also shows  MEs and MUEs for
the BH6 test set,\cite{ae6bh6} which has been introduced as a computationally less intensive, but statistically representative subset of HT/NHTBH38.  
However, we do not present a detailed discussion on BH6 here,  but  stress that
errors in BH6 essentially follow the trends found for HT/NHTBH38.

One of the main findings of this work is the astonishingly good performance of the conventional \RPA\
scheme for activation energies.
To be more specific, \RPA\ performs significantly better 
for the transfer of hydrogen atoms than for  reactions involving heavier atoms.
For HTBH38, the ME obtained using \RPA\ amounts to
 \unit[$-0.8$]{kJ/mol} and the associated MUE amounts to \unit[$7.1$]{kJ/mol}.
These error margins are similar to those
 of some of the range-separated, generalized KS-DFT functionals like {\it e.g.}~\LRPBE.\cite{vydrov:jcp:06}
The latter performs very well for chemical reaction barriers (see also Sec.~\ref{sec:compare}).
However, for \RPA, the MUE increases by more than 50\% when elements heavier than H, like {\it e.g.}~F or Cl, are transferred.
The MUE in NHTBH38 obtained using \RPA\ amounts to \unit[$12.1$]{kJ/mol} 
together with a rather distinct underestimation of the barriers by \unit[$-10.5$]{kJ/mol} (ME).

On a  relative scale, the MURE for HT reactions obtained using \RPA\ (see Fig.~\ref{fig:barriers})
amounts to approximately 20\%, but increases to a value approximately twice as large for NHT reactions
[panel (b)].
Note that reaction 7 in NHTBH38 has a barrier height of only \unit[$-1.42$]{kJ/mol}.
For this reaction MUREs are extraordinarily large leading
to an increase, which is seven to eight times as large as the corresponding value in HT reactions.
The statistics would be drastically biased by such a case being very likely compensated by
significantly extending the test set.
Therefore, we decided to exclude reaction number 7 from the test set when calculating the MURE in NHTBH38.
Both \RPAh\ and \RPASE\ show a strong underestimation of barriers with maximal errors  as large as
 \unit[$50$]{kJ/mol}.
As mentioned above, the reason for this behavior has been attributed to
small HOMO-LUMO differences found for some of the transition state structures, which
are not correctly described by the simple \RPASE\ scheme.
Indeed, the renormalization of \cSE\ alleviates the problem,
and the corresponding ME and MUE  in HTBH38 obtained using \RPARSE\ are improved by almost 60\% compared to \RPASE.
%
Note that  numerical results given in Tab.~\ref{tab:res} nicely reflect the trend induced by incorporation of SE effects in the correlation
energy contribution, {\it i.e.} it partially takes care of the lack of self-consistency in the EX@PBE energy.
However, the rSE corrects for the strong underestimation of barriers seen in \RPAh\ and \RPASE, but qualitatively reflects the same trend
compared to \RPA.

The performance of \SOSEX\ for barrier heights has already been tested by Paier {\it et al.} for 
the BH6 test set.\cite{paier:jcp:10}
This work  extends the findings of Ref.~\onlinecite{paier:jcp:10} by  discriminating
 HT and NHT reactions.
\SOSEX\ is less accurate for HT barriers than \RPA\ as indicated by an MUE of about  \unit[$22$]{kJ/mol} compared to \unit[$7$]{kJ/mol}.
Quantitatively, \SOSEX\ on average { overestimates} barrier heights for HTBH38 by the aforementioned \unit[$22$]{kJ/mol}.
This  is in perfect agreement with the errors found for the BH6 test set.\cite{paier:jcp:10}
On the other hand, \SOSEX\ performs substantially better for NHT barrier heights,
where ME and MUE are found to be close to the ones obtained using \RPA.
On average, \SOSEX\ { overestimates} NHT barriers by approximately \unit[$13$]{kJ/mol},
whereas \RPA\ underestimates them by \unit[$11$]{kJ/mol}.
As shown in Fig.~\ref{fig:barriers}, the MURE in NHT barriers obtained using \SOSEX\ amounts to 34\% [panel (b) in Fig.~\ref{fig:barriers}]
slightly outperforming \RPA\ by approximately 3\%.

Incorporation of  SE effects into \SOSEX\ in the hybrid fashion, {\it i.e.}~\SOSEXh,
 leads to very different results when applied to HT and NHT reactions, respectively.
\SOSEXh\  improves  HT reaction barrier heights, 
whereas NHT barrier heights deteriorate appreciably compared to \SOSEX,
ending up with an overall { underestimation} of  barrier heights.

The situation becomes noticeably better, for both HT and NHT barrier heights, upon 
combination of explicitly computed renormalized SE with \SOSEX.  
Barrier heights obtained using \SOSEXRSE\ are of similar quality as ``conventional'' \RPA,
although the unsigned error in NHT test set is slightly larger.
\SOSEXRSE\ 
overestimates HT barriers by approximately \unit[$3.6$]{kJ/mol}, but
reduces the ME in NHT barriers (ME = \unit[$-6.3$]{kJ/mol}) compared to \RPA.

To summarize this section, SOSEX and \cRSE\ tend to overestimate and underestimate 
reaction barrier heights, respectively. 
Thus it appears  advantageous  to combine the correction schemes in order 
to achieve a partial error compensation.
This mechanism works most efficiently for HT reactions and somewhat less so for NHT reactions.  
Taking the 
excellent performance of \SOSEXRSE\ 
for binding energies (see previous Section) into account, this functional offers the most 
balanced description in terms of binding energies as well as activation energies.

\begin{table*}[htb]
\renewcommand{\arraystretch}{1.1}
\setlength{\tabcolsep}{9pt}
\parbox{12cm}{
\caption{\label{tab:reac_ene}
Mean  errors and mean unsigned errors [kJ/mol] in
the reaction energies obtained using calculated barrier heights of the HTBH38/04 hydrogen transfer
as well as  NHTBH38/04 non-hydrogen transfer barrier heights.
}
}
\begin{tabular}{lrr@{\hspace{12mm}}rr}
\hline
\hline
              & \multicolumn{2}{l}{HTBH38} & \multicolumn{2}{c}{NHTBH38}\\
    Method    &   ME   &   MUE  &    ME  &   MUE   \\
\hline
\RPA          &  $-$3.2   & 18.2                 &  $-$7.8                & 9.7      \\
\RPAh         &  2.2      & 12.3                 &    $-$1.6               &  14.4       \\
\RPASE       &  $-$3.0     & 16.9                 &   9.4                & 24.6        \\
\RPARSE      &  $-$2.9     &  17.0                &  $-$1.2                 &  11.8        \\
\SOSEX    &  2.7      & 4.6                  &   $-$18.4                &  20.5       \\
\SOSEXh   &  2.8      & 4.1                  &    $-$12.2               &  15.7       \\
\SOSEXRSE&   3.0     &   4.4                &       $-$11.9            &  15.5        \\
\hline
\hline
\end{tabular}
\end{table*}

\subsection{Reaction energies in HTBH38 and NHTBH38}

\begin{figure}
\includegraphics[width=.99\columnwidth,clip=true]{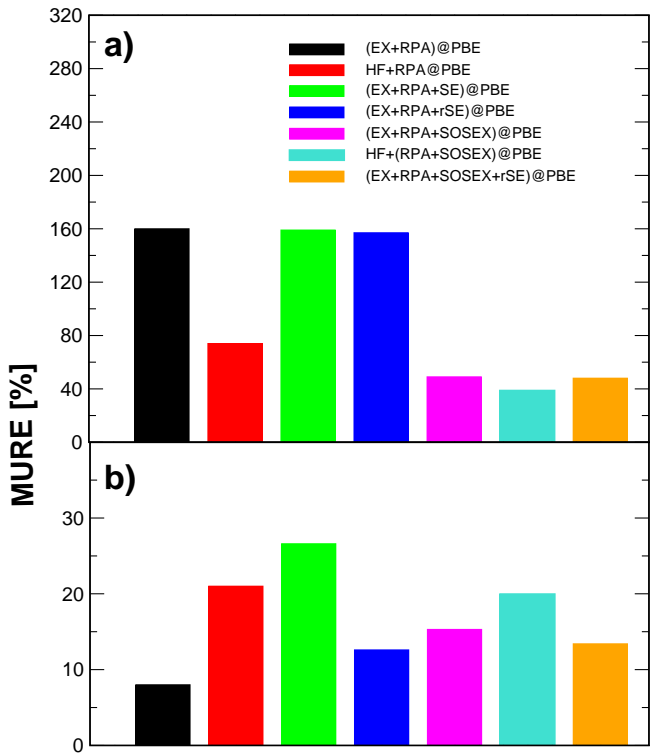}
\caption{\label{fig:reactionene} 
Panel a) shows mean unsigned relative errors (MURE) in HT reaction energies of HTBH38.
Panel b) shows MURE for the reaction energies of NHTBH38. Energies use a cc-pV(T,Q)Z extrapolation and
the frozen core approximation to the correlation energies. Color code of legends follows Fig.~\ref{fig:barriers}.}
\end{figure}

As shown in Fig.~\ref{fig:activation}, knowing both forward ($\text{V}^{\neq}_{\text{f}}$) and reverse ($\text{V}^{\neq}_{\text{r}}$) reaction barrier heights,  corresponding
reaction energies $\Delta\text{E}_{\text{}}$ are readily calculated using
\begin{equation}
\label{eq:reac}
\Delta\text{E}_{\text{}} = \text{V}^{\neq}_{\text{f}} - \text{V}^{\neq}_{\text{r}}.
\end{equation}
Note that 17 out of the 38 reactions contained in HTBH38 lead to a nonzero $\Delta\text{E}$,
whereas  NHTBH38 comprises 13 reactions with a forward barrier different from the reverse barrier.
The corresponding
MEs and MUEs of the RPA-based functionals are compiled in Tab.~\ref{tab:reac_ene},
and the MUREs are depicted in Fig.~\ref{fig:reactionene}

Similar to the trends found for atomization energies, HT reaction energies 
are significantly improved upon inclusion of (\ccSOSEX)@PBE\ to \RPA\ as reflected in the MUEs.
For \RPA\, the MUE in HT reactions amounts to \unit[$18.2$]{kJ/mol} and drops down to \unit[$4.6$]{kJ/mol}
 employing \SOSEX.
Hence, it appears that eliminating
 the one-electron self-correlation error contained in \cRPA@PBE is
beneficial for HT reaction energies.
This is not entirely surprising, since the aforementioned
error will be largest for breaking and creating  covalent hydrogen bonds.
For reactions involving heavier atoms, as exemplified by the reaction energies in NHTBH38, the correction due
to (\ccSOSEX)@PBE appears to perform less favorably.
This can be seen by inspection of Fig.~\ref{fig:reactionene} presenting MUREs in HT  [panel (a)] as well as NHT reaction energies [panel (b)].
For \RPA\ the MUE in NHTBH38 amounts to \unit[$9.7$]{kJ/mol}, which is rather low, whereas for
NHT reaction energies obtained using \SOSEX, the MUE increases to \unit[$20.5$]{kJ/mol}.

Concerning  effects due to \cSE@PBE\ and \cRSE@PBE\ to \RPA, no significant improvement
of HT reaction energies over \RPA\ has been found.
The MEs and MUEs given in Tab.~\ref{tab:reac_ene} for \RPASE\ (ME = $-3$ kJ/mol; MUE = $16.9$ kJ/mol)
and \RPARSE\ (ME = $-2.9$ kJ/mol; MUE = $17$ kJ/mol) are essentially unaltered compared to \RPA.
In contrast to HT, the rSE correction helps to improve the NHTBH38 reaction energies and alleviates the overestimation found for simple \RPASE\  drastically (ME = $-1.2$ kJ/mol compared to 9.4 kJ/mol).
The associated  MUE as well as MURE decrease by approximately 50\%.

We now turn to a discussion of results obtained using the 
``hybrid variants'', which employ the  HF energy as
the reference energy on the EX level.
Specifically for \RPA, HT reaction energies are substantially improved
upon replacement of EX@PBE through HF.
As can be seen from Tab.~\ref{tab:reac_ene}, the MUE is reduced by approximately \unit[$6$]{kJ/mol}, which
translate into an improvement of the MURE by approximately 50\%.
HT reaction energies obtained using \SOSEX, which are fairly accurate, are hardly 
affected by  changing to the corresponding hybrid scheme.
Employing \SOSEXh, however, the MUE in NHT reaction energies is reduced by \unit[$5$]{kJ/mol}.
In addition, the ME amounts to \unit[$-12$]{kJ/mol},  which compares very favorably to the ME of 
\unit[$-18$]{kJ/mol} obtained using \SOSEX.
In terms of performance, the combined scheme \SOSEXRSE\ is on par with \SOSEXh\ for both HTBH38 and NHTBH38 
reaction energies.
\begin{table*}[htb]
\renewcommand{\arraystretch}{1.2}
\setlength{\tabcolsep}{8pt}
\parbox{12.2cm}{
\caption{\label{tab:compare}
Comparing the three best-performing functionals presented in Tab.~\ref{tab:res} to widely used semilocal and HF/DFT hybrid functionals. 
Mean unsigned errors in individual test sets are given in kJ/mol.
}}
\begin{threeparttable}
\begin{tabular}{lrrrr}
\hline
\hline
 Method     &  G2-1              &    BH6        &  HTBH38         & NHTBH38          \\
\hline
(EX+\cRPA)@PBE    & 42.8$^{\rm a}$            &    7.5$^{\rm a}$         & 7.1$^{\rm \phantom{a}}$            &  12.1$^{\rm \phantom{a}}$ \\
HF+(\cSOSEX)@PBE  & \phantom{1}13.0$^{\rm \phantom{a}}$ & 9.2$^{\rm \phantom{a}}$           &13.8$^{\rm \phantom{a}}$           &  25.5$^{\rm \phantom{a}}$  \\
(EX+\cSOSEX+\cRSE)@PBE   &  13.9$^{\rm \phantom{a}}$   &  3.7$^{\rm \phantom{a}}$           & 5.4$^{\rm \phantom{a}}$           &  17.6$^{\rm \phantom{a}}$   \\
\hline
PBE                &   36.0$^{\rm b}$            &  38.9$^{\rm c}$          & 39.0$^{\rm d}$             &  33.9$^{\rm d}$       \\
BLYP               &   19.7$^{\rm b}$            &  32.6$^{\rm c}$          & 31.5$^{\rm d}$              &  36.4$^{\rm d}$        \\
PBE0               &   14.6$^{\rm b}$            &  19.2$^{\rm c}$          & 17.7$^{\rm d}$              &  14.1$^{\rm d}$        \\
B3LYP              &   10.0$^{\rm b}$            &  19.7$^{\rm c}$          & 17.7$^{\rm d}$              &  18.2$^{\rm d}$         \\      
LC-$\omega$PBE              &   15.6$^{\rm e}$            &           & 5.4$^{\rm e}$              &  10.0$^{\rm e}$         \\      
\hline
\hline
\end{tabular}
\begin{tablenotes}
\item[a] See Ref.~\onlinecite{paier:jcp:10}. Note that differences in the MUE of G2-1 are due to the different values for the experimental
dissociation energies (see Ref.~\onlinecite{feller:jcp:99}).
\item[b] Ref.~\onlinecite{PBE1PBE}
\item[c] Ref.~\onlinecite{yang:jcp:10}
\item[d] Ref.~\onlinecite{NHTBH38}
\item[e] Ref.~\onlinecite{vydrov:jcp:06}. Note that the MUE given here for G2 refers to G2-2 comprising 148 molecules. The MUE for G2-1 will be lower.
\end{tablenotes}
\end{threeparttable}
\end{table*}
\SOSEXRSE\ has two apparent favorable features: (i) it
substantially improves HT reaction energies obtained using \RPA, and (ii) it performs approximately
similarly well for {\em all} of the test sets investigated in this work.
In other words, the overall variation in  error margins for atomization energies, barrier heights, and reaction 
energies is smallest for \SOSEXRSE\ lending the functional robustness.
Among the functionals discussed in this work, \RPA\  performs best for
 NHT reaction energies.
Nevertheless, \SOSEXRSE\ performs only slightly worse, but given
the better HT reaction barrier heights and the significantly better reaction energies in HTBH38, \SOSEXRSE\ is among the RPA-based functionals tested in this work, the functional of broadest applicability.

\subsection{Comparing RPA to semilocal and hybrid functionals\label{sec:compare}}

To close the discussion on the performance of the RPA- and RPA+SOSEX-based functionals,
we briefly compare  molecular atomization and activation energies to results obtained using commonly applied semilocal as well
as HF/DFT hybrid functionals.
To render direct comparisons easier, Table~\ref{tab:compare} repeats MUEs for G2-1, BH6, HTBH38, and NHTBH38 for  three of the
RPA-based functionals, which perform best, namely (EX+\cRPA)@PBE, HF+(\cSOSEX)@PBE, and (EX+\cSOSEX+\cRSE)@PBE.
The above mentioned statistical errors are compared to PBE-GGA, BLYP-GGA\cite{becke:pra:88,lee:prb:88} as well as the PBE0\cite{PBE0,PBE1PBE} and 
B3LYP\cite{B3LYP} global hybrid functionals.
In addition, we also present results obtained using the above mentioned LC-$\omega$PBE range-separated hybrid functional.\cite{vydrov:jcp:06}
\LRPBE\ mixes a fraction of EX at the long-range of the Coulomb interaction (for definitions, see Ref.~\onlinecite{vydrov:jcp:06}), but
uses only one parameter (0.40 bohr$^{-1}$) for defining a universal range separation.
It is remarkable that LC-$\omega$PBE describes reaction barriers and atomization energies extremely accurately   representing a landmark among hybrids for thermochemistry and kinetics.
Admittedly, for extended systems admixture of EX on the long-range is detrimental and leads to {\it e.g.}~strongly overestimated band gaps.\cite{gerber:jcp:07}

Returning to RPA, activation energies obtained using 
\RPA\  are {\it de facto } on par with LC-$\omega$PBE (Tab.~\ref{tab:compare}).
Trends for GGA and global hybrid functionals like PBE0 or B3LYP
 are rather general, hence other GGA-type and global hybrid functionals  perform
quite similarly (for other functionals, see {\it e.g.}~Ref.~\onlinecite{yang:jcp:10}).
Although, HF+(\cSOSEX)@PBE does not perform as well as \RPA\ for activation energies of non-hydrogen transfer reactions (corresponding MUE is almost twice as large), it performs certainly better than PBE and BLYP.
 HF+(\cSOSEX)@PBE is only slightly outperformed by B3LYP for
the aforementioned activation energies in NHTBH38.
According to this synopsis, (EX+\cSOSEX+\cRSE)@PBE certainly shows the most balanced description of molecular binding and barrier heights.
It performs similarly well as hybrid functionals in terms of atomization energies, 
 outperforms
both GGA and hybrid functionals in terms of hydrogen-transfer barrier heights, and performs equivalently well for non-hydrogen
barrier heights as  aforementioned hybrids do.

Although, this work is not devoted to weak, van-der-Waals-type of interactions, it should be emphasized that all of the RPA-based
functionals presented here do include them seamlessly as already mentioned in the introduction.
It is well known that neither GGA nor hybrid functionals do show the correct van der Waals asymptote.

\section{Conclusions\label{sec:conc}}
In summary, we have reported an extensive assessment of several exact-exchange based
post-KS density functionals involving RPA correlation energies and beyond. 
Correlation energies have been assessed for solids as well as for small molecules.
Specifically we calculated atomization energies of solids and molecules using \RPA, \SOSEX\ as well
as \RPAh\ and \SOSEXh, where the latter approach gives binding energies improved by approximately 50\% compared
to ``conventional'' \RPA.
Furthermore, we investigated the performance of individual functionals
for chemical reaction barrier heights or activation energies
employing large and well established test sets.
Generally, we found that it is rather difficult to systematically improve on \RPA\ reaction barrier heights, although
modest improvements using \SOSEXRSE\ were achieved for HT barriers.
Importantly, the favorable impact of the correlation energy contribution stemming from SE effects
 on binding energies does not translate to reaction barriers.
This has been explained by  divergent correlation energy contributions in systems with small HOMO-LUMO gaps.
Therefore, application of ``\cSE'' to systems with small one-electron band gaps  is not possible, but a renormalization of ``\cSE'' helps to alleviate the problem.
Surprisingly, \RPA\ yields reaction energies of high accuracy for reactions
involving non-hydrogen atoms.
Good and robust performance of a novel RPA-based functional
\SOSEXRSE\ is a central finding of this work.
It improves on binding or atomization energies compared to \RPA, improves on HT barrier heights as well
as reaction energies.

\section*{Acknowledgements}
This work was supported by the Austrian Fonds zur F{\"o}rderung der 
wissenschaftlichen Forschung (FWF) within SFB ViCom (F41).
The work at Rice University was supported by the US Department of Energy 
(Grant No. DE-FG02-09ER16053) and The Welch Foundation (C-0036).


\begin{thebibliography}{100}

\bibitem{Bohm/Pines:1953}
D.~Bohm and D.~Pines,
\newblock Phys. Rev. {\bf 92}, 609 (1953).

\bibitem{Gell-Mann/Brueckner:1957}
M.~Gell-Mann and K.~A. Brueckner,
\newblock Phys. Rev. {\bf 106}, 364 (1957).

\bibitem{Furche:2001}
F.~Furche,
\newblock Phys. Rev. B {\bf 64}, 195120 (2001).

\bibitem{Fuchs/Gonze:2002}
M.~Fuchs and X.~Gonze,
\newblock Phys. Rev. B {\bf 65}, 235109 (2002).

\bibitem{furche:jcp:05}
F.~Furche and T.~{Van Voorhis},
\newblock J. Chem. Phys. {\bf 122}, 164106 (2005).

\bibitem{Scuseria/Henderson/Sorensen:2008}
G.~E. Scuseria, T.~M. Henderson, and D.~C. Sorensen,
\newblock J. Chem. Phys. {\bf 129}, 231101 (2008).

\bibitem{Janesko/Henderson/Scuseria:2009}
B.~G. Janesko, T.~M. Henderson, and G.~E. Scuseria,
\newblock J. Chem. Phys. {\bf 130}, 081105 (2009).

\bibitem{Toulouse/etal:2009}
{J. Toulouse, I.~C. Gerber, G. Jansen, A. Savin, J.~G. \'Angy\'an},
\newblock Phys.\ Rev.\ Lett. {\bf 102}, 096404 (2009).

\bibitem{paier:jcp:10}
J.~Paier, B.~G. Janesko, T.~M. Henderson, G.~E. Scuseria, A. Gr\"uneis, and G. Kresse,
\newblock J. Chem. Phys. {\bf 132}, 094103 (2010),
\newblock Erratum: {\it ibid.}~{\bf 133}, 179902 (2010).

\bibitem{Marini/Gonzalez/Rubio:2006}
A.~Marini, P.~Garc{\'i}a-Gonz{\'a}lez, and A.~Rubio,
\newblock Phys. Rev. Lett {\bf 96}, 136404 (2006).

\bibitem{Jiang/Engel:2007}
H.~Jiang and E.~Engel,
\newblock J. Chem. Phys {\bf 127}, 184108 (2007).

\bibitem{Harl/Kresse:2008}
J.~Harl and G.~Kresse,
\newblock Phys. Rev. B {\bf 77}, 045136 (2008).

\bibitem{Harl/Kresse:2009}
J.~Harl and G.~Kresse,
\newblock Phys. Rev. Lett. {\bf 103}, 056401 (2009).

\bibitem{Lu/Li/Rocca/Galli:2009}
D.~Lu, Y.~Li, D.~Rocca, and G.~Galli,
\newblock Phys. Rev. Lett. {\bf 102}, 206411 (2009).

\bibitem{Dobson/Wang:1999}
J.~F. Dobson and J.~Wang,
\newblock Phys. Rev. Lett {\bf 82}, 2123 (1999).

\bibitem{Rohlfing/Bredow:2008}
M.~Rohlfing and T.~Bredow,
\newblock Phys. Rev. Lett {\bf 101}, 266106 (2008).

\bibitem{Ren/Rinke/Scheffler:2009}
X.~Ren, P.~Rinke, and M.~Scheffler,
\newblock Phys.\ Rev.\ B {\bf 80}, 045402 (2009).

\bibitem{Schimka/etal:2010}
{L. Schimka, J. Harl, A. Stroppa, A. Gr{\"u}neis, M. Marsman, F. Mittendorfer, and G. Kresse },
\newblock Nature Materials {\bf 9}, 741 (2010).

\bibitem{Zhu/etal:2010}
{W. Zhu, J. Toulouse, A. Savin, and J.~G. \'Angy\'an},
\newblock J. Chem. Phys. {\bf 132}, 244108 (2010).

\bibitem{Lebegue/etal:2010}
S.~Leb\`egue, J. Harl, T. Gould, J.~G. \'Angy\'an, G. Kresse,  and J.~F. Dobson,
\newblock Phys. Rev. Lett. {\bf 105}, 196401 (2010).

\bibitem{eshuis:jcp:2010}
H.~Eshuis, J.~Yarkony, and F.~Furche,
\newblock J. Chem. Phys. {\bf 132}, 234114 (2010).

\bibitem{Ismail-Beigi:2010}
S.~Ismail-Beigi,
\newblock Phys. Rev. B {\bf 81}, 195126 (2010).

\bibitem{goltl:jcp:2011}
F.~G{\"o}ltl and J.~Hafner,
\newblock J. Chem. Phys. {\bf 134}, 064102 (2011).

\bibitem{eshuis:jpcl:2011}
H.~Eshuis and F.~Furche,
\newblock J. Phys. Chem. Lett. {\bf 2}, 983 (2011).

\bibitem{ren:prl:11}
X.~Ren, A.~Tkatchenko, P.~Rinke, and M.~Scheffler,
\newblock Phys. Rev. Lett. {\bf 106}, 153003 (2011).

\bibitem{Hesselmann/Goerling:2011}
A.~He\ss{}elmann and A.~G\"orling,
\newblock Phys. Rev. Lett. {\bf 106}, 093001 (2011).

\bibitem{Hesselmann/Goerling:2011_2}
A.~He\ss{}elmann and A.~G\"orling,
\newblock Mol. Phys. {\bf 109}, 2473 (2011).


\bibitem{ruzsinszky:jcp:2011}
A.~Ruzsinszky, J.~P. Perdew, and G.~I. Csonka,
\newblock J. Chem. Phys. {\bf 134}, 114110 (2011).

\bibitem{hesselmann:jcp:11}
A.~He{\ss}elmann,
\newblock J. Chem. Phys. {\bf 134}, 204107 (2011).

\bibitem{klopper:cpl:11}
W.~Klopper, A.~M. Teale, S.~Coriani, T.~B. Pedersen, and T.~Helgaker,
\newblock Chem. Phys. Lett. {\bf 510}, 147 (2011).

\bibitem{Dobson:1994}
J.~F. Dobson,
\newblock in {\em \textit{Topics in Condensed Matter Physics}}, edited by M.~P.
  Das, Nova, New York, 1994.

\bibitem{Moller/Plesset:1934}
C.~M{\o}ller and M.~S. Plesset,
\newblock Phys. Rev. {\bf 46}, 618 (1934).

\bibitem{BHH}
A.~D. Becke,
\newblock J. Chem. Phys. {\bf 98}, 1372 (1993).

\bibitem{B3}
A.~D. Becke,
\newblock J. Chem. Phys. {\bf 98}, 5648 (1993).

\bibitem{PBE14}
J.~P. Perdew, M.~Ernzerhof, and K.~Burke,
\newblock J. Chem. Phys. {\bf 105}, 9982 (1996).






\bibitem{freeman:prb:77}
D.~L. Freeman,
\newblock Phys. Rev. B {\bf 15}, 5512 (1977).

\bibitem{grueneis:jcp:09}
A.~Gr{\"u}neis, M.~Marsman, J.~Harl, L.~Schimka, and G.~Kresse,
\newblock J. Chem. Phys. {\bf 131}, 154115 (2009).

\bibitem{harl:prb:10}
J.~Harl, L.~Schimka, and G.~Kresse,
\newblock Phys. Rev. B {\bf 81}, 115126 (2010).

\bibitem{henderson:molphys:10}
T.~M. Henderson and G.~E. Scuseria,
\newblock Mol. Phys. {\bf 108}, 2511 (2010).


\bibitem{Manyes}
A.~Ruzsinszky, J.~P. Perdew, G.~I. Csonka, O.~A. Vydrov, and G.~E. Scuseria,
\newblock J. Chem. Phys. {\bf 125}, 194112 (2006).

\bibitem{MCY_on_SIE}
P.~Mori-S\'anchez and A.~J. Cohen,
\newblock J. Chem. Phys. {\bf 125}, 201102 (2006).


\bibitem{hu:prb:86}
C.~D. Hu and D.~C. Langreth,
\newblock Phys. Rev. B {\bf 33}, 943 (1986).

\bibitem{kresse:elba}
G.~Kresse and A.~Gr{\"u}neis,
\newblock unpublished results, October 3 (2008),
\newblock presented at the XIV ESCMQC, Isola d'Elba, Italy.


\bibitem{curtiss:jcp:97}
L.~A. Curtiss, K.~Raghavachari, P.~Redfern, and J.~Pople,
\newblock J. Chem. Phys. {\bf 106}, 1063 (1997).

\bibitem{pople:jcp:89}
J.~A. Pople, M.~{Head-Gordon}, D.~J. Fox, K.~Raghavachari, and L.~A. Curtiss,
\newblock J. Chem. Phys. {\bf 90}, 5622 (1989).

\bibitem{curtiss:jcp:89}
L.~A. Curtiss, C.~Jones, G.~W. Trucks, K.~Raghavachari, and J.~A. Pople,
\newblock J. Chem. Phys. {\bf 93}, 2537 (1989).

\bibitem{curtiss:jcp:98}
L.~A. Curtiss, P.~C. Redfern, K.~Raghavachari, and J.~A. Pople,
\newblock J. Chem. Phys. {\bf 109}, 42 (1998).

\bibitem{BH42}
Y.~Zhao, B.~J. Lynch, and D.~G. Truhlar,
\newblock J. Phys. Chem. A {\bf 108}, 2715 (2004).

\bibitem{NHTBH38}
Y.~Zhao, N.~Gonz\'ales-Garc\'ia, and D.~G. Truhlar,
\newblock J. Phys. Chem. A {\bf 109}, 2012 (2005),
\newblock {\bf{110}}, 4942(E) (2006).

\bibitem{janesko:jcp:08}
B.~G. Janesko and G.~E. Scuseria,
\newblock J. Chem. Phys. {\bf 128}, 244112 (2008).


\bibitem{GKS}
A.~Seidl, A.~G\"orling, P.~Vogl, J.~A. Majewski, and M.~Levy,
\newblock Phys. Rev. B {\bf 53}, 3764 (1996).


\bibitem{sharp:pr:53}
R.~T. Sharp and G.~K. Horton,
\newblock Phys. Rev. {\bf 90}, 317 (1953).

\bibitem{talman:pra:76}
J.~D. Talman and W.~F. Shadwick,
\newblock Phys. Rev. A {\bf 14}, 36 (1976).

\bibitem{casida:pra:95}
M.~E. Casida,
\newblock Phys. Rev. A {\bf 51}, 2005 (1995).

\bibitem{szabo}
A.~Szabo and N.~S. Ostlund,
\newblock {\em Modern Quantum Chemistry},
\newblock Dover, Mineola, New York, 1$^{\rm st}$ edition, 1996.

\bibitem{DreizlerGross}
R.~M. Dreizler and E.~K.~U. Gross,
\newblock {\em Density Functional Theory},
\newblock Plenum Press, New York, 1995.

\bibitem{Godby/Schlueter/Sham:1986}
R.~W. Godby, M.~Schl\"uter, and L.~J. Sham,
\newblock Phys.\ Rev.\ Lett. {\bf 56}, 2415 (1986).

\bibitem{Godby/Schluter/Sham:1988}
R.~W. Godby, M.~Schl\"uter, and L.~J. Sham,
\newblock Phys. Rev. B {\bf 37}, 10159 (1988).

\bibitem{Kotani:1998}
T.~Kotani,
\newblock J. Phys.: Condens. Matter {\bf 10}, 9241 (1998).

\bibitem{Gruening/Marini/Rubio:2006}
M.~Gr\"uning, A.~Marini, and A.~Rubio,
\newblock Phys.\ Rev.\ B {\bf R74}, 161103 (2006).

\bibitem{Hellgren/Barth:2007}
M.~Hellgren and U.~{von Barth},
\newblock Phys. Rev. B {\bf 76}, 075107 (2007).

\bibitem{Hellgren/Rohr/Gross:2012}
M.~Hellgren, D.~R. Rohr, and E.~K.~U. Gross,
\newblock J. Chem. Phys. {\bf 136}, 034106 (2012).

\bibitem{langreth:ssc:75}
D.~C. Langreth and J.~P. Perdew,
\newblock Solid. State. Commun. {\bf 17}, 1425 (1975).

\bibitem{gunnarsson:prb:76}
O.~Gunnarsson and B.~I. Lundqvist,
\newblock Phys. Rev. B {\bf 13}, 4274 (1976).

\bibitem{langreth:prb:77}
D.~C. Langreth and J.~P. Perdew,
\newblock Phys. Rev. B {\bf 15}, 2884 (1977).

\bibitem{gross:prl:kohn}
E.~K.~U. Gross and W.~Kohn,
\newblock Phys. Rev. Lett. {\bf 55}, 2850 (1985).

\bibitem{RubioReining}
G.~Onida, L.~Reining, and A.~Rubio,
\newblock Rev. Mod. Phys. {\bf 74}, 601 (2002).

\bibitem{harris}
F.~E. Harris, H.~J. Monkhorst, and D.~L. Freeman,
\newblock {\em Algebraic and Diagrammatic Methods in Many-Fermion Theory},
\newblock Oxford University Press, New York, Oxford, 1992.


\bibitem{brandow:rmp:67}
B.~H. Brandow,
\newblock Rev. Mod. Phys. {\bf 39}, 771 (1967).

\bibitem{bartlett:rmp:07}
R.~J. Bartlett and M.~Musia{\l},
\newblock Rev. Mod. Phys. {\bf 79}, 291 (2007).

\bibitem{kummel:prep:78}
H.~K{\"u}mmel, K.~H. L{\"u}hrmann, and J.~G. Zabolitzky,
\newblock Phys. Rep. {\bf 36}, 1 (1978).

\bibitem{luhrmann:prb:78}
R.~F. Bishop and K.~H. L{\"u}hrmann,
\newblock Phys. Rev. B {\bf 17}, 3757 (1978).

\bibitem{luhrmann:prb:82}
R.~F. Bishop and K.~H. L{\"u}hrmann,
\newblock Phys. Rev. B {\bf 26}, 5523 (1982).

\bibitem{cizek:jcp:66}
J.~\v{C}\'i\v{z}ek,
\newblock J. Chem. Phys. {\bf 45}, 4256 (1966).

\bibitem{cizek:adcp:69}
J.~\v{C}\'i\v{z}ek,
\newblock Adv. Chem. Phys. {\bf 14}, 35 (1969).

\bibitem{paldus:pra:72}
J.~Paldus, J.~\v{C}\'i\v{z}ek, and I.~Shavitt,
\newblock Phys. Rev. A {\bf 5}, 50 (1972).

\bibitem{bartlett:ijqc:78}
R.~J. Bartlett and G.~D. {Purvis~III},
\newblock Int. J. Quantum Chem. {\bf 14}, 561 (1978).

\bibitem{bartlett:arpc:81}
R.~J. Bartlett,
\newblock Ann. Rev. Phys. Chem. {\bf 32}, 359 (1981).


\bibitem{hedin:pr:65}
L.~Hedin,
\newblock Phys. Rev. {\bf 139}, A796 (1965).

\bibitem{nozieres:nuocim:58}
P.~Nozi{\`e}res and D. Pines,
\newblock Nuovo Cimento {\bf 9}, 470 (1958).


\bibitem{monkhorst:prl:73}
H.~J. Monkhorst and J.~Oddershede,
\newblock Phys. Rev. Lett. {\bf 30}, 797 (1973).

\bibitem{jansen:jcp:10}
G. Jansen, {R.-F.}~Liu, and J.~G. \'Angy\'an
\newblock J. Chem. Phys. {\bf 133}, 154106 (2010).

\bibitem{gorlinglevy:prb:93}
A. G{\"o}rling and M.~Levy
\newblock Phys. Rev. B {\bf 47}, 131105 (1993).

\bibitem{raffenetti:tca:92}
R.~C. Raffenetti, K. Ruedenberg, C.~L. Janssen, and H.~F. Schaefer,
\newblock Theor. Chim. Acta {\bf 86}, 149 (1992).

\bibitem{ruedenberg:ijqc:76}
K. Ruedenberg, L.~M. Cheung, and S.~T. Elbert,
\newblock Int. J. Quantum. Chem. {\bf 16}, 1069 (1976).

\bibitem{scuseria:cpl:87}
G.~E. Scuseria and H.~F. Schaefer,
\newblock Chem. Phys. Lett. {\bf 142}, 354 (1987).

\bibitem{PBE}
J.~P. Perdew, K.~Burke, and M.~Ernzerhof,
\newblock Phys. Rev. Lett. {\bf 77}, 3865 (1996),
\newblock {(E)}~{\it ibid.} \textbf{78}, 1396 (1997).

\bibitem{eshuis/bates/furche:tca:12}
H.~Eshuis, J.~E.~Bates, and F.~Furche,
\newblock Theor. Chem. Acc. {\bf 131}, 1084 (2012).


\bibitem{kresse:prb:93}
G.~Kresse and J.~Hafner,
\newblock Phys. Rev. B {\bf 48}, 13115 (1993).

\bibitem{kresse:96}
G.~Kresse and J.~Furthm{\"u}ller,
\newblock Phys. Rev. B {\bf 54}, 11169 (1996).

\bibitem{kresse:96_2}
G.~Kresse and J.~Furthm{\"u}ller,
\newblock Comput. Mater. Sci. {\bf 6}, 15 (1996).

\bibitem{gdv-g1}
Gaussian Development Version, Revision G.01, M.~J. Frisch, G.~W. Trucks, H. B. Schlegel{\em et. al.},
  Gaussian, Inc., Wallingford CT, 2007.

\bibitem{blum:cpc:09}
V.~Blum, R. Gehrke, F. Hanke, P. Havu, V. Havu, X. Ren, K. Reuter, and M. Scheffler
\newblock Comput. Phys. Commun. {\bf 180}, 2175 (2009).

\bibitem{ren:prb:sub}
X.~Ren, P.~Rinke, V.~Blum, J.~Wieferink, A.~Tkatchenko, A.~Sanfilippo, K.~Reuter, and M.~Scheffler,
\newblock to be published.

\bibitem{ae6bh6}
B.~J. Lynch and D.~G. Truhlar,
\newblock J. Phys. Chem. A {\bf 107}, 8996 (2003),
\newblock {\bf{108}}, 1460(E) (2004).

\bibitem{kutzelnigg:jcp:92}
W.~Kutzelnigg and J.~D. {Morgan III},
\newblock J. Chem. Phys. {\bf 96}, 4484 (1992).

\bibitem{helgaker:jcp:97}
T.~Helgaker, W.~Klopper, H.~Koch, and J.~Noga,
\newblock J. Chem. Phys. {\bf 106}, 9639 (1997).

\bibitem{halkier:cpl:98}
A.~Halkier, T. Helgaker, P. J{\o}rgensen, W. Klopper, H. Koch, J. Olsen, and A.~K. Wilson,
\newblock Chem. Phys. Lett. {\bf 286}, 243 (1998).

\bibitem{dunning:jcp:89}
T.~H. {Dunning,~Jr.},
\newblock J. Chem. Phys. {\bf 90}, 1007 (1989).

\bibitem{woon:jcp:93}
D.~E. Woon and T.~H. {Dunning, Jr.},
\newblock J. Chem. Phys. {\bf 98}, 1358 (1993).

\bibitem{boys:molphys:70}
S.~F. Boys and F.~Bernardi,
\newblock Mol. Phys. {\bf 19}, 553 (1970).

\bibitem{grueneis:jcp:10}
A.~Gr{\"u}neis, M.~Marsman, and G.~Kresse,
\newblock J. Chem. Phys. {\bf 133}, 074107 (2010).

\bibitem{staroverov:prb:04}
V.~N. Staroverov, G.~E. Scuseria, J.~Tao, and J.~P. Perdew,
\newblock Phys. Rev. B {\bf 69}, 075102 (2004).

\bibitem{madelung:sdh:04}
O.~Madelung,
\newblock {\em Semiconductors: Data Handbook},
\newblock Springer, Berlin, 3rd edition, 2004.

\bibitem{trampert:book:1998}
A.~Trampert, O.~Brandt, and K.~Ploog,
\newblock {\em Semiconductors and Semimetals}, volume~50, chapter Crystal
  Structure of Group III Nitrides,
\newblock Academic, San Diego, 1998,
\newblock edited by J.~I. Pankove and T.~D. Moustakas.


\bibitem{smith:japplc:68}
D.~K. Smith and H.~R. Leider,
\newblock { J. Appl. Cryst.} {\bf 1}, 246 (1968).


\bibitem{gruneis:jctc:11}
A. Gr{\"u}neis, G.~H. Booth, M. Marsman, J. Spencer, A. Alavi, and G.~Kresse,
\newblock J. Chem. Theory Comput. {\bf 7}, 2780 (2011).

\bibitem{feller:jcp:99}
D.~Feller and K.~A. Peterson,
\newblock J. Chem. Phys. {\bf 110}, 8384 (1999).

\bibitem{schimka:jcp:11}
L.~Schimka, J.~Harl, and G.~Kresse,
\newblock J. Chem. Phys. {\bf 134}, 024116 (2011).

\bibitem{vydrov:jcp:06}
O.~Vydrov and G.~E. Scuseria,
\newblock J. Chem. Phys. {\bf 125}, 234109 (2006).

\bibitem{PBE1PBE}
M.~Ernzerhof and G.~E. Scuseria,
\newblock J. Chem. Phys. {\bf 110}, 5029 (1999).

\bibitem{yang:jcp:10}
K.~Yang, J.~Zheng, Y.~Zhao, and D.~G. Truhlar,
\newblock J. Chem. Phys. {\bf 132}, 164117 (2010).

\bibitem{becke:pra:88}
A.~D. Becke,
\newblock Phys. Rev. A {\bf 38}, 3098 (1988).

\bibitem{lee:prb:88}
C.~Lee, W.~Yang, and R.~G. Parr,
\newblock Phys. Rev. B {\bf 37}, 785 (1988).

\bibitem{PBE0}
C.~Adamo and V.~Barone,
\newblock J. Chem. Phys. {\bf 110}, 6158 (1999).

\bibitem{B3LYP}
P.~J. Stephens, F.~J. Devlin, C.~F. Chabalowski, and M.~J. Frisch,
\newblock J. Phys. Chem. {\bf 98}, 11623 (1994).

\bibitem{gerber:jcp:07}
I.~C. Gerber, J.~G. \'Angy\'an, M. Marsman, and G.~Kresse,
\newblock J. Chem. Phys. {\bf 127}, 054101 (2007).


\end{thebibliography}

\end{document}